\documentclass[aps, pre, preprint]{revtex4-1}
\pdfoutput=1
\usepackage[utf8]{inputenc}
\usepackage[english]{babel}

\usepackage{graphicx}
\usepackage{amsmath}
\usepackage{amssymb}
\usepackage{float}
\usepackage{bm}
\usepackage{natbib}
\usepackage{color}

\newcommand{\e}{\mathrm{e}}
\newcommand{\overbar}[1]{\mkern 1.5mu\overline{\mkern-1.5mu#1\mkern-1.5mu}\mkern 1.5mu}
\newcommand{\degree}{\ensuremath{^\circ}}

\begin{document}
\title{Stochastic efficiency of thermodiffusion: an extended local equilibrium approach}
\author{Jean-François Derivaux and Yannick De Decker}
\affiliation{Nonlinear Physical Chemistry Unit, Service de Chimie Physique et Biologie Th\'eorique\\ and\\ Center for Nonlinear Phenomena and Complex
Systems (CENOLI)\\
Universit\'e libre de Bruxelles (ULB), Campus Plaine, C.P. 231. B-1050 Brussels, Belgium}
\begin{abstract}
The recently established connection between stochastic thermodynamics and fluctuating hydrodynamics is applied to a study of efficiencies in the coupled transport of heat and matter on a small scale. 
A stochastic model for a mesoscopic cell connected to two macroscopic reservoirs of heat and particles is developed, based on fluctuating hydrodynamics.
Within this approach, the fluctuating separation and thermodynamic efficiencies are defined. The conditions required to observe bimodal distributions of
these efficiencies are determined, and the evolution of these distributions is investigated in the large-size and in the long-time limits.
The results obtained  are not restricted to thermodiffusion and can be generalized to systems where efficiency
is defined as a ratio of stochastic state variables or dissipation rates.
\end{abstract}

\maketitle

\section{Introduction}

Thermodynamics was originally developed in the 19th century to understand conversion of energy in macroscopic 
systems such as pumps or steam engines. Over the years, along with progress made in the analysis and in the manipulation of micro- and 
nanoscales devices, the interest in extending thermodynamics to small systems has been growing. At such scales fluctuations of the state variables 
cannot be  overlooked and need to be incorporated explicitly. The numerous efforts made in this direction  
have led to the emergence of a new field collectively referred to as stochastic thermodynamics 
\cite{schlogl_stochastic_1980,jiu-li_stability_1984,gaspard_fluctuation_2004,sekimoto_stochastic_2010,seifert_stochastic_2012,van_den_broeck_stochastic_2014,ciliberto_fluctuations_2010}.

One recurrent and important question in stochastic thermodynamics is how to quantify the dynamics of relevant thermodynamic quantities, starting 
from the mesoscopic-level rules of stochastic processes. Most approaches so far have been based on statistical definitions of the thermodynamic quantities 
(such as the Gibbs-Shannon expression for entropy) combined with evolution equations for the probability distributions of state variables. 
Recently, we proposed a novel framework which rests on an extension of classical non-equilibrium thermodynamics, and thus of the local equilibrium hypothesis, to derive stochastic 
differential equations describing the spatiotemporal evolution of thermodynamic quantities of interest \cite{de_decker_extended_2015}.
This formalism was already applied to chemical reactions \cite{de_decker_stochastic_2016} and  
Brownian motion \cite{nicolis_stochastic_2017}.
In the present paper we use it  to analyze the fluctuating thermodynamics of a model for thermodiffusion in small systems.

Thermodiffusion consists in a coupling of transport phenomena, whereby gradients of temperature can induce transport of matter and gradients 
of concentrations can lead to heat flow. Because of this coupling, it becomes for instance possible to transport mass from a diluted reservoir to a 
more concentrated one.
A system where such phenomena occur can be seen 
as an engine due to its ability to convert heat in work done against a gradient.
Depending on the aim of the device, different relevant measures of the efficiency of thermodiffusion can be defined. The two most common efficiencies 
are the separation efficiency, which is a ratio between the gradients of mass and temperature, and  
the thermodynamic efficiency, which quantifies the relative dissipation of energy needed to induce thermodiffusion.

Both these efficiencies share the common property of being ratios of thermodynamic quantities. 
Various ratios of this type were previously studied in theoretical and experimental works. 
Ratios of work and heat were for example observed to display bimodal distributions for quantum dots and for Brownian motors 
\cite{proesmans_stochastic_2015,martinez_brownian_2016}. 
A theoretical analysis of the thermodynamic efficiency  was carried out later in \cite{polettini_efficiency_2015}. Bimodality 
was there interpreted as a sign  that the system can work in different distinct regimes, giving rise to two peaks in 
the distribution of efficiency. 

In the present work, we use the extended local equilibrium hypothesis to study the statistical properties of  efficiencies 
in a simple model for thermodiffusion in compartmentalized systems. We first present the model and discuss its properties in the macroscopic 
(mean-field) limit (Section \ref{sec:deterministic_model}). In Section  \ref{sec:stochastic_model} we show how fluctuations can be incorporated
thanks to the aforementioned extension of the local equilibrium assumption. 
Stochastic versions of the separation and thermodynamic efficiencies are then introduced and analyzed in Section \ref{sec:stochastic_efficiencies}. 
The distributions of these quantities are studied 
numerically, and analytically when possible. Special emphasis is put on the parametric conditions leading to the emergence of bimodal 
probability distributions. 
We also study in detail how the efficiencies behave in the thermodynamic limit, i.e., for long times and/or large volumes. 
Interestingly, the way the thermodynamic limit is taken can have a strong impact on the distributions and these effects are not 
the same for the separation and for the thermodynamic efficiencies. These results are summarized 
and put in perspective in Section \ref{Conclusions}, where we also discuss possible extensions of our work.

\section{Deterministic Model}
\label{sec:deterministic_model}

We consider in this work an ideal and non-reactive dilute binary solution, with no bulk velocity and no external force.   
In the framework of a spatially continuous description, the evolution equations for the local  
temperature $T$ of the solution and for the local mass  density $\rho$ of the solute are 
given by the divergence of fluxes: 
\begin{align} 
\rho_{tot}\, c_v\, \frac{\partial T}{\partial t} &= - \text{div} \,\textbf{J}_q  \label{eq:eq_evol_T_cont} \\ 
\frac{\partial \rho}{\partial t} &= - \text{div} \,  \textbf{J}_1, \label{eq:eq_evol_rho_cont} 
\end{align}
with $c_v$ and $\rho_{tot}$ being the specific heat at constant volume and the mass density of the solution, respectively. 
In the linear regime of non-equilibrium thermodynamics, the heat flux $\textbf{J}_q$ and the flux of matter $\textbf{J}_1$ are both 
linear combinations of the thermodynamic forces involved: 
\begin{align}
\textbf{J}_q &= - L_{qq} \, \frac{ \nabla T }{T^2} - L_{q1} \, \frac{\nabla_T  \mu}{T} \label{eq:continuous_heat_flux} \\
\textbf{J}_1 &= -  L_{1q} \, \frac{\nabla  T }{T^2} - L_{11} \, \frac{ \nabla_T \mu}{T}  \label{eq:continuous_matter_flux}.
\end{align}
In these equations, $\mu$ is the chemical potential per unit mass and $\nabla_T$ 
stands for gradients taken at constant temperature.  
The coefficients before the thermodynamics forces are the Onsager coefficients and the off-diagonal terms obey the Onsager reciprocity relation 
$L_{q1} = L_{1q}$.
The fluxes are usually expressed by relating the Onsager coefficients to experimentally accessible  quantities \cite{groot_non-equilibrium_2011},  
\begin{align}
\textbf{J}_{q} &= - \kappa \, \nabla T - \rho\, \mu^{\prime} \, T \, D_T\,  \nabla \rho \\
\textbf{J}_1 &= - \rho \,  \,  D_T \nabla T - D \, \nabla \rho,
\end{align}
where 
\begin{eqnarray}
 \kappa = \frac{L_{qq}}{T^2}
 \label{eq:thermal_conductivity}
\end{eqnarray}
is the thermal conductivity, 
\begin{eqnarray}
 D_T = \frac{L_{q1}}{\rho\, T^2} 
 \label{eq:thermal_diffusion_coeff}
\end{eqnarray}
is the thermal diffusion coefficient and 
\begin{eqnarray}
 D = \frac{L_{11}}{T}\,  \mu^{\prime}
 \label{eq:diffusion_coeff}
\end{eqnarray}
is the diffusion coefficient.
Note that in this formulation information is required on   the partial derivative of the chemical potential of the solute, 
$\mu^{\prime} = \left( \frac{\partial \mu}{\partial \rho} \right)_{T,p,\rho_{solv}}$, where $p$ is the pressure and 
$\rho_{solv}$ is the mass density of the solvent.  
Since we consider an ideal and dilute solution, the chemical 
potential takes the form 
\begin{equation}
\label{eq:chemical_pot}
\mu = \mu \degree(T,p) +  R_s\, T \, \ln \rho,
\end{equation}
where $\mu \degree$ is a standard chemical potential and $R_s$ is the specific gas constant (the gas constant divided by  the molar mass of the solute).  
The expressions for the fluxes thus finally read:
\begin{align}
\label{eq:equations_fluxes_Jq}
\textbf{J}_{q} &= - \kappa \, \nabla T -  R_s\, T^2 \, D_T\, \nabla \rho \\
\textbf{J}_1 &= - \rho \, D_T \, \nabla T - D\,  \nabla \rho\label{eq:equations_fluxes_J1}.
\end{align}

\begin{figure}[h!]
\centering
\includegraphics[scale = 1.0]{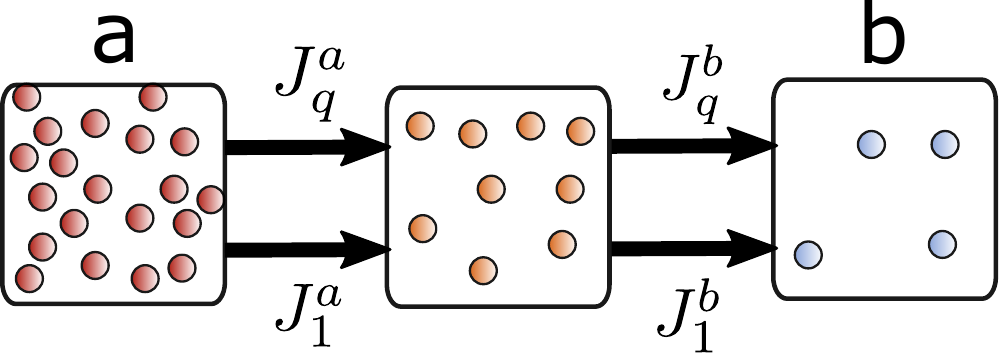}
\caption{Scheme of the model studied in this paper. 
A mesoscopic system is connected to two reservoirs denoted by $a$ and $b$. 
Each reservoir can exchange heat and matter with the central system.}
\label{fig:scheme_model}
\end{figure}

Our objective in this work is to analyze the properties of stochastic efficiencies  
for the simplest model where thermodiffusion can occur. 
We consider to this end a homogeneous  (well-stirred) system of volume $\Omega$, which is 
connected to two homogeneous reservoirs from which heat and matter can flow, as depicted in Fig. \ref{fig:scheme_model}. 
Each reservoir plays both the role of a chemostat and a thermostat.
This model could be representative of  mesoscopic channels or pores, where 
thermodiffusion is already used to sort particles \cite{lervik_sorting_2014} or for DNA sequencing \cite{he_thermophoretic_2013}.
The evolution laws for the internal energy $U$ and for the mass $M$ of solute in the central subsystem are given by 
integrating the fluxes of heat and matter flowing across the surfaces separating the reservoirs and the system,   
\begin{align} \label{dU}
\frac{dU}{dt} =  - \sum_k \int_{S^k} \textbf{J}_q^k \cdot \textbf{dS}^k \equiv \sum_k F^k_q \\ \label{dM}
\frac{dM}{dt} = - \sum_k \int_{S^k} \textbf{J}_1^k \cdot \textbf{dS}^k  \equiv  \sum_k F^k_1,  
\end{align}
where $k=(a,b)$ refers to the reservoir under consideration and $S^k$ is the surface separating the central system from a reservoir $k$. 
In what follows, we will consider that these surfaces all have the same area $S$. 
To evaluate these evolution equations explicitly, 
we suppose that all gradients can be  written as differences
between  values  in the subsystems involved divided by the length of the central system (see \cite{groot_non-equilibrium_2011} Chapter XV),
\begin{eqnarray}
 \nabla T = \frac{\Delta T}{l} \quad ; \quad \nabla \rho = \frac{\Delta \rho}{l}.
\end{eqnarray}
Inserting these expressions for the gradients in the fluxes \eqref{eq:equations_fluxes_Jq} and  \eqref{eq:equations_fluxes_J1} 
leads to the following expressions for the rates appearing in \eqref{dU}-\eqref{dM}:
\begin{eqnarray} \label{eq:heat_fluxesa}
F_q^a &=&  - \kappa^{\prime}\,  (T-T_a) -  R_s\,  T^2\,  D_T^{\prime} \, (\rho - \rho_a) \\ \label{eq:heat_fluxesb}
F_q^b &=& - \kappa^{\prime}\, (T-T_b) -  R_s \,  T^2\,  D_T^{\prime} \,  (\rho -\rho_b)  \\  \label{eq:diffusive_fluxesa}
F_1^a &=& - D^{\prime} (\rho - \rho_a) - \rho D_T^{\prime} (T-T_a) \\ \label{eq:diffusive_fluxesb}
F_1^{b} &=& - D^{\prime} (\rho-\rho_b) - \rho D_T^{\prime} (T-T_b).  
\end{eqnarray}
Here, $T_a$ ($T_b$)  and $\rho_a$ ($\rho_b$) are the constant temperature and density of reservoir $a$ ($b$).  
These equations include the rescaled transport coefficients 
\begin{eqnarray}
\kappa^{\prime} = \frac{S}{l}\, \kappa \quad ; \quad ; D_T^{\prime} = \frac{S}{l}\, D_T \quad ; \quad D^{\prime} = \frac{S}{l}\, D.  
\end{eqnarray}
Notice that in a way similar to what is found in the spatially continuous description, the rates can be expressed 
as linear combinations of (here discrete) thermodynamic forces \cite{groot_non-equilibrium_2011}:
\begin{align}
F_q^k = -L_{qq}^{\prime} \frac{\left( T - T_k \right)}{T^2} - L_{q1}^{\prime} \frac{\mu - \mu_k}{T} \\
F_1^k =  - L_{q1}^{\prime} \frac{\left( T - T_k \right)}{T^2} - L_{11}^{\prime} \frac{\mu - \mu_k}{T},
\end{align}
in which one can find rescaled version of the continuous Onsager coefficients
\begin{eqnarray}
L_{qq}^{\prime} = \frac{S}{l}\, L_{qq} \quad ; \quad L_{q1} = \frac{S}{l}\, L_{q1} \quad ; \quad L_{11}^{\prime} = \frac{S}{l}\, L_{11}. 
\label{eq:Onsager_coeff_discrete}  
\end{eqnarray}

Inserting the rates \eqref{eq:heat_fluxesa}-\eqref{eq:diffusive_fluxesb} 
in eqs. \eqref{dU}-\eqref{dM}, and using the relations $dU= \Omega\, \rho_{tot}\, c_v\, dT$ and $dM=\Omega\, d\rho$, the evolution equations for the temperature 
and for the mass density of solute in the central system are finally obtained: 
\begin{align} \label{dT}
\frac{dT}{dt} &= \gamma \, \lambda \, (T_a + T_b - 2\, T) + \gamma\, \alpha D_T \, T^2\,  (\rho_a + \rho_b- 2\, \rho) \\ \label{drho}
\frac{d\rho}{dt} &=  \gamma\, D\,  (\rho_a + \rho_b - 2\, \rho) + \gamma\,  D_T\,  \rho\,  (T_a + T_b- 2\, T).
\end{align}
We introduced in these equations the geometric factor 
\begin{eqnarray}
\gamma = \frac{S}{\Omega\, l} 
\end{eqnarray}
which is the inverse of a specific area and the quantities 
\begin{eqnarray}
 \lambda = \frac{\kappa}{\rho_{tot}\, c_v}
\end{eqnarray}
and
\begin{eqnarray}
 \alpha = \frac{ R_s}{\rho_{tot}\,  c_v}. 
\end{eqnarray}
Because the binary mixture is ideal and dilute, 
the transport coefficients $\kappa$, $D$, $D_T$ and the density of the solution $\rho_{tot}$ do not depend on $\rho$. 
Furthermore, we will assume for simplicity that these parameters are independent of temperature. 
Non-linearities in eqs. \eqref{dT}-\eqref{drho} 
thus solely arise from switching from Onsager coefficients to transport coefficients, as well as from the derivative of the chemical potential \eqref{eq:chemical_pot}. 
More complex situations could be encountered by considering non-constant transport coefficients, 
but we prefer to focus here on the simplest non-linear case.

Eqs. \eqref{dT}-\eqref{drho}, which form our model,  
can be seen as the discrete analogues of the continuous equations \eqref{eq:eq_evol_T_cont} and \eqref{eq:eq_evol_rho_cont}. 
They form an autonomous 2-variable dynamical system which can admit up to three steady-state solutions. One of these solutions is 
\begin{equation} \label{steady_1}
\left(T_{s,1},\rho_{s,1} \right) = \left(\frac{T_a + T_b}{2},  \frac{\rho_a + \rho_b}{2}\right)
\end{equation}
and depends on external constraints only. Since it includes the equilibrium state as a special case, we 
will name it the \textit{thermodynamic branch}. 
The two other solutions depend both on reservoir-related quantities and on kinetic parameters:
\begin{gather} \label{steady_2}
T_{s,2-3} = \frac{ \lambda \, \boldsymbol{\mp} \sqrt{\lambda \,(\alpha\, (\rho_a+\rho_b) (2 \, D-D_T \,(T_a+T_b))+\lambda)}}{\alpha \, D_T (\rho_a+\rho_b)}\\ \label{steady_3}
\rho_{s,2-3} = \frac{D \, \left(\alpha\, (\rho_a + \rho_b)\left[2\, D -D_T\, (T_a+T_b)\right] + 2 \lambda  \boldsymbol{\pm} 2
\sqrt{ \lambda\, (\alpha\, (\rho_a+\rho_b) \left[2\, D-D_T \, (T_a+T_b)\right]+ \lambda)}\right)}{\alpha \left[2\, D- D_T\, (T_a+T_b)\right]^2}. 
\end{gather}
Depending on the values of the different parameters involved, the system can thus present one or three 
physically relevant solutions.

\begin{figure}[H]
\centering
\includegraphics[scale = 0.70]{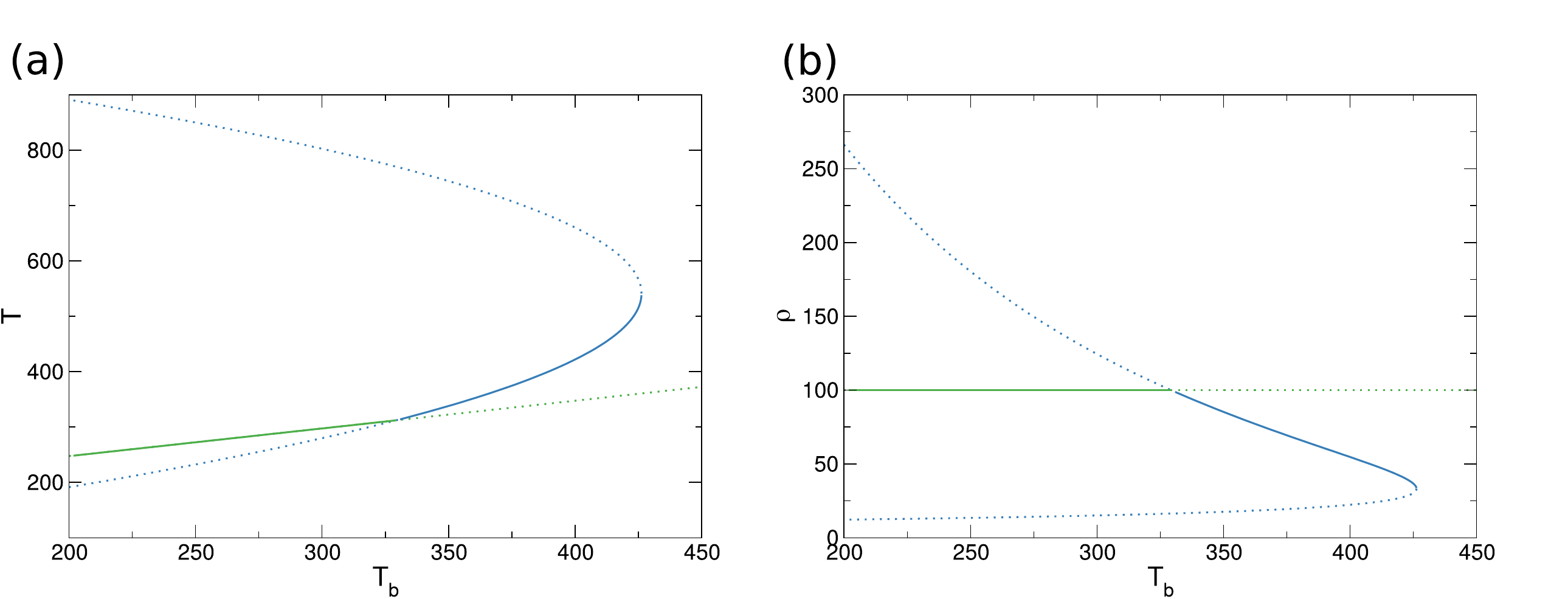}
\caption{Bifurcation diagram of the temperature (a) and of the density (b) as a function of 
the control parameter $T_b$, as obtained from eqs. \eqref{steady_1}-\eqref{steady_3}. Solid lines represent the stable steady states and dotted lines the unstable states. 
The green color stands for the thermodynamic branch and blue for the other solutions. Parameters: $T_a  =295$ K,
$\rho_a = 110$ kg$\cdot$m$^{-3}$, $\rho_b = \, 90$ kg$\cdot$m$^{-3}$, $\rho_{tot} = 1000$ kg$\cdot$m$^{-3}$, $\kappa = 9\times10^{-6}$ J$\cdot$m$^{-1} \cdot$K$^{-1}\cdot$s$^{-1}$, 
$D = 9\times 10^{-11}$ m$^{2}\cdot$s$^{-1}$, $D_T =10^{-12}$ m$^{2}\cdot$K$^{-1}\cdot$s$^{-1}$, $  R_s = 83.14$ 
J$\cdot$K$^{-1}\cdot$kg$^{-1}$, $c_v =4200$ J$\cdot$K$^{-1}\cdot$kg$^{-1}$, $\gamma =  10^{16}$ m$^{-2}$.}
\label{fig:set_bd}
\end{figure}

A typical example of bifurcation diagram corresponding to this system is depicted in 
Fig. \ref{fig:set_bd}, where the temperature $T_b$ of reservoir $b$ is varied. 
At low $T_b$, the only stable solution is the thermodynamic branch. 
Increasing $T_b$, a transcritical bifurcation takes place  
at which the thermodynamic
branch loses stability in favor of one of the two solutions \eqref{steady_2}-\eqref{steady_3}.
The chosen set of parameters is representative of typical aqueous solutions and we thus expect that this transition could be observed experimentally. 
Note however that the transcritical bifurcation only appears for moderate values of $T_b$ when the thermal conductivity is low: 
this would correspond to  a system whose walls are made of a moderately insulating material.  
 As $T_b$ is further increased, these two additional solutions 
merge at a limit point bifurcation whose coordinates are given by 
\begin{equation}
T_b = \frac{\lambda + \alpha \, \left(\rho_a+\rho_b \right) \, (2 \, D- D_T\,  T_a)}{D_T\, \alpha \, \left(\rho_a+ \rho_b\right)}
\end{equation}
At this limit point, the degenerate steady-state values of $T$ and $\rho$ take simple forms: 
\begin{align}
{T}_{s,2} &= {T}_{s,3} = \frac{\lambda}{D_T \, \alpha \, (\rho_a + \rho_b)} \\
{\rho}_{s,2} &= {\rho}_{s,3} = \frac{(\rho_a + \rho_b)^2\,  D\,  \alpha}{\lambda}.
\end{align} 
This limit point marks the end of the domain of validity of the model developed here. Beyond this point, 
the system does not admit any stable steady state and both the temperature and the density ``explode''. 
The absence of physically meaningful states in this parametric region is more than probably due 
to the fact that for such large temperatures, the linear forms we used to estimate the fluxes are no longer relevant. 
A similar effect is also found for too large density gradients. We will thus restrict ourselves to cases where the limit point bifurcation has 
not been crossed. 

\section{Stochastic model}
\label{sec:stochastic_model}

We consider in this work mesoscopic systems where fluctuations are not negligible. In order to take them into account, stochastic 
differential equations for the mass of solute and for the internal energy of the system need to be derived. To do so,  
we will here rely on the recently introduced extended local equilibrium framework \cite{de_decker_extended_2015}. In this approach, 
the local equilibrium hypothesis at the basis of classical non-equilibrium thermodynamics is supposed to 
hold even when the evolution equations for the state variables are stochastic. 
The statistical properties of the entropy production associated with simple diffusion processes \cite{de_decker_extended_2015}, chemical reactions 
\cite{de_decker_stochastic_2016} and Brownian motion \cite{nicolis_stochastic_2017} were already investigated in this framework.

Due to the presence of fluctuations, the heat and the mass fluxes contain an additional noise term and are fluctuating quantities,  
\begin{eqnarray}
 \tilde{\textbf{J}}_q &=& \textbf{J}_q + {\bm{\xi}}_q\\
 \tilde{\textbf{J}}_1 &=& \textbf{J}_1 + \bm{\xi}_1,
\end{eqnarray}
where the first terms in the right-hand sides are the systematic fluxes given by eqs. \eqref{eq:equations_fluxes_Jq}-\eqref{eq:equations_fluxes_J1} 
and the noise terms are denoted by the vectors $\bm{\xi}$. 
The spectrum of fluctuations for the state variables can become intricate in non-equilibrium  states, exhibiting 
for example long-range correlations. A large amount of literature has been devoted to elucidate the causes of such features. 
They were related to the presence of multiplicative noise, whose amplitude depends on the spatial position 
(\textit{inhomogeneous noise}) \cite{tremblay_fluctuations_1981}, on advection terms\cite{kirkpatrick_light_1982, ortiz_de_zarate_physical_2004} 
or on state variables \cite{pagonabarraga_long-range_1994}. Since the system introduced in this paper is homogeneous, well-stirred and with constant transport 
coefficients, we consider that  Landau's classical 
fluctuating hydrodynamics \cite{lifshitz_statistical_1980, zarate_hydrodynamic_2006} holds. 
Consequently, 
the additive noise terms correspond to zero-mean Gaussian white noises whose amplitudes obey the fluctuation-dissipation
relations \cite{foch_stochastic_1971, zarate_hydrodynamic_2006}
\begin{align}
\overbar{\xi^i_q\left( t \right) \xi^j_q \left(t^{\prime} \right)} &= 2\,  k_B\, \kappa \,  T_m^2\ 
\, \delta\left( t- t^{\prime} \right) \, \delta\left( \bm{r}- \bm{r}^{\prime} \right) \, \delta_{ij}  \label{eq:amplitude_noise_continuous_heat}  \\
\overbar{\xi^i_1\left( t \right) \xi^j_1 \left(t^{\prime} \right)} &= 2\, k_B\, \frac{ D \, \rho_m}{R_s}\, 
\delta\left( t- t^{\prime} \right) \, \delta\left(  \bm{r}- \bm{r}^{\prime}  \right) \, \delta_{ij}  \\
\overbar{\xi^i_q\left( t \right) \xi^j_1 \left(t^{\prime} \right)} &=  2\, k_B\, D_T\, \rho_m \, 
T_m^2\,  \delta\left( t- t^{\prime} \right)\, \delta\left(  \bm{r}- \bm{r}^{\prime}  \right) \, \delta_{ij}  \label{eq:amplitude_noise_continuous_heat_matter},
\end{align}
in which  $i$ and $j$ represent the different contributions in space and $k_B$ is Boltzmann's constant. The lines over $\xi_q$, $\xi_1$ represent averaging over the noise.  
$T_m$ and $\rho_m$ are the temperature and the mass density evaluated along the macroscopic (deterministic) trajectories computed by solving the deterministic 
equations \eqref{dT} - \eqref{drho}. 
The stochastic evolution equations for $U$ and $M$ are now given by 
\begin{align} \label{dU_stoch}
\frac{dU}{dt} =  - \sum_k \int_{S^k} \left( \textbf{J}_q + \boldsymbol{\xi}_q  \right) \cdot \textbf{dS}^k \equiv \sum_k \left(F^k_q + f^k_q\right) \\ \label{dM_stoch}
\frac{dM}{dt} = - \sum_k \int_{S^k} \left( \textbf{J}_1 + \boldsymbol{\xi}_1  \right) \cdot \textbf{dS}^k  \equiv  \sum_k \left(F^k_1 + f^k_1\right) 
\end{align}
The extended local equilibrium hypothesis has the consequence that the ``equation of state'' $dU=  \Omega\, \rho_{tot}\, c_v\, dT$ holds even in the presence of 
fluctuations. Consequently, starting now from eqs. \eqref{dU_stoch}-\eqref{dM_stoch}  
the evolution equations of the intensive state variables read 
\begin{align}
\frac{dT}{dt} &=  \gamma\,  \lambda\, (T_a + T_b - 2\, T) +  \gamma\, \alpha\,  D_T\, T^2 \, (\rho_a + \rho_b- 2\, \rho) + R_T(t) \equiv V_T+R_T(t)\label{eq:SDE_T} \\
\frac{d\rho}{dt} &=   \gamma\, D\, (\rho_a + \rho_b - 2\, \rho) +  \gamma\, D_T\, \rho\, (T_a + T_b- 2\, T) + R_{\rho}(t) \equiv V_\rho+R_\rho(t). \label{eq:SDE_C}
\end{align}
$R_T(t)$ and $R_{\rho}(t)$ are zero-mean Gaussian white noises obtained by combination of the fluctuating contributions to the fluxes:
\begin{align}
R_T(t) &= \frac{f^a_q + f^b_q}{\Omega\, \rho_{tot}\,  c_v}  \label{eq:R_T}\\
R_{\rho}(t) &= \frac{f^a_1 + f^b_1}{\Omega} \label{eq:R_C}.
\end{align} 
They are both zero-mean Gaussian variables, whose covariances read
\begin{align}
\overline{R_T(t)\, R_T(t')} &=   \frac{1}{\Omega} \, \left[4\, k_B\,  \frac{\gamma\, \lambda }{\rho_{tot}\,  c_v}\,  T_m^2\right] 
\, \delta\left( t- t^{\prime} \right) \equiv R_{TT}\, \delta\left( t- t^{\prime} \right)\\
\overline{R_{\rho}(t)\, R_{\rho}(t')} &=  \frac{1}{\Omega} \left[4\, k_B\, \frac{ \gamma\, D}{  R_s} \, \rho_m \right]\, 
\delta\left( t- t^{\prime} \right)  \equiv R_{\rho \rho}\, \delta\left( t- t^{\prime} \right)\\
\overline{R_T(t)\, R_\rho(t')} & =  \frac{1}{\Omega}\,  \left[ 4\, k_B\, \frac{\gamma \, D_T}{\rho_{tot}\,  c_v}\, \rho_m \, T_m^2\right]\, 
\delta\left( t- t^{\prime} \right)  \equiv R_{T \rho}\, \delta\left( t- t^{\prime} \right).
\end{align}

The noise terms being entirely characterized, the analytical distribution of deviations of state variables from their mean values, 
$\delta X_i = X_i - \overbar{X_i}$, can be computed
in the weak-noise limit. 
 To do so, we expand the evolution equations
around the means:
\begin{align}
\frac{d}{dt} \ \delta T &=  \left( \frac{\partial V_T}{\partial T}\right)_{\overbar{T}, \overbar{\rho}} \delta T +
\left( \frac{\partial V_{T}}{\partial \rho}\right)_{\overbar{T}, \overbar{\rho}} \delta \rho + \ldots + R_T(t) \label{eq:Langevin_fluctuations_DT}\\
\frac{d}{dt} \ \delta \rho &= \left( \frac{\partial V_{\rho}}{\partial T}\right)_{\overbar{T}, \overbar{\rho}} \delta T + 
\left( \frac{\partial V_{\rho}}{\partial T}\right)_{\overbar{T}, \overbar{\rho}} \delta \rho + \ldots + R_{\rho}(t)\label{eq:Langevin_fluctuations_DC}
\end{align}
Keeping the first order terms and introducing the matrix $\beta$ with elements 
\begin{equation}
\beta_{i,j} = - \left( \frac{\partial V_i}{\partial X_j}\right)_{\{\overbar{X_i}\}}
\end{equation} 
the evolution of deviations can be written in the form of coupled Ornstein-Uhlenbeck processes, 
\begin{equation}
\frac{d}{dt} \, \delta X_i = \sum_j - \beta_{i,j}\,  \delta X_j + R_i(t) \text{,}
\end{equation}
which obey a bivariate Gaussian distribution:
\begin{equation}
P\left(\boldsymbol{\delta X}\right) = \frac{1}{2\pi \sqrt{\text{det}(C)}} \, \e^{-\frac{1}{2}\left( \boldsymbol{\delta X}^\mathsf{T} C^{-1} \boldsymbol{\delta X}  \right)}
\end{equation}
where $\boldsymbol{\delta X}$ is the vector containing the fluctuations of state variables, $\boldsymbol{\delta X} = \left( \delta T, \delta \rho \right)$, and $C$ is the 
equal-time covariance matrix of the state variables:
\begin{equation}C = 
\begin{pmatrix}
\overline{\delta T^2(t)} & \overline{\delta T(t)\, \delta \rho(t)} \\
\overline{\delta T(t)\, \delta \rho(t)} & \overline{\delta \rho^2(t)}
\end{pmatrix}
\equiv 
\begin{pmatrix}
C_{TT} & C_{T\rho} \\
C_{T\rho} & C_{\rho \rho}
\end{pmatrix}
\end{equation}
The latter can be  computed from the matrix of rates derivatives $\beta$ \cite{risken_fokker-planck_2013}:
\begin{equation}
C\beta^\mathsf{T} + \beta \, C = Q
\label{eq:relation_covariance_matrices}
\end{equation}
where 
\begin{equation}Q = 
\begin{pmatrix}
R_{TT} & R_{T\rho} \\
R_{T\rho} & R_{\rho \rho}
\end{pmatrix}
\end{equation} is the covariance matrix of the noise terms $R_T, R_\rho$.

\section{Efficiencies}
\label{sec:stochastic_efficiencies}

The efficiency of a thermodiffusive process can be quantified in several different ways. 
The \textit{separation efficiency} $\chi_k$ is the ratio between 
the gradient of density and the gradient of temperature with respect to a given reservoir $k$:
\begin{equation} \label{sep_eff}
\chi_k = \frac{\rho - \rho_k}{T-T_k}, \qquad k = a,b.
\end{equation}
Evaluating this quantity is especially relevant when thermodiffusion is used to concentrate or to deplete a solute with the help
of temperature gradients.
Alternatively, as mentioned before  the system can also be seen  as a sort of engine where a usually non-spontaneous transport 
process can arise thanks to coupling with another transport phenomenon.
In this context, a relevant measure is the \textit{thermodynamic efficiency} $\eta_k$, 
\begin{equation}\label{thermo_eff}
\eta_k = - \frac{\sigma_-^k}{\sigma_+^k}
\end{equation}
where $\sigma_{-}^k<0$ is the entropy production of the  non-spontaneous process and $\sigma_+^k>0$ 
the entropy production of the  spontaneous process, both related to the reservoir $k$. The thermodynamic efficiency 
quantifies the (relative) dissipative cost associated with the emergence of the desired transport
phenomenon. 
In what  follows, we will be interested in the study of the statistical properties of those two efficiencies in a non-equilibrium steady state in which the state variables $T$ and $\rho$ fluctuate around their stationary macroscopic values 
$T_m$ and $\rho_m$, as given by eq. (29) on the 
thermodynamic branch and eqs. (30)-(31) on the post-bifurcation branch.
Under these conditions the probability of the state variables is expected to relax towards its stationary distribution.  This relaxation
is typically very rapid for the parameter values used in the manuscript. We thus rapidly reach a state where both the mean and the variance of
$T$ and $\rho$ are stationary as well.
\subsection{Stochastic separation efficiency}
\label{subsec:stochastic_separation_efficiency}

Because we include fluctuations in our description of the system, the separation efficiency  \eqref{sep_eff} is the ratio 
of two state variables obeying stochastic differential equations. In the weak-noise limit, the 
evolution equations \eqref{eq:SDE_T}-\eqref{eq:SDE_C}  define two coupled Ornstein-Uhlenbeck processes and the variables are thus expected 
to be distributed according to Gaussian distributions. It has been shown in various cases that the  
probability distribution of similar ratios can exhibit bimodality \cite{polettini_efficiency_2015,rana_single-particle_2014}. 
We investigate here whether the same phenomenon can be observed for the distribution of the separation efficiency. More precisely, we 
will be interested in the influence of external control parameters (such as 
the temperature or the density of the reservoirs), of the size of the system and of the duration of observation on the emergence of 
bimodality.

\subsubsection{Probability distribution of $\chi_k$}

Inasmuch as $\chi_k$ can be seen as the ratio of two random Gaussian variables, its probability distribution $P(\chi_k)$ is given by \cite{polettini_efficiency_2015}
\begin{equation}
P(\chi_k) = \frac{1}{\pi \sqrt{\text{det}(C)} f(\chi_k)}\, \e^{-h} \left[ 1 + \sqrt{\pi}\, j(\chi_k) \, \e^{ j(\chi_k)^2}\,  \text{erf} \left( j(\chi_k) \right)\right]
\label{eq:distrib_chik}
\end{equation}
where
\begin{gather}
f(\chi_k) = \frac{1}{\text{det}(C)} \left( C_{TT}\, \chi_k^2 - 2 \, C_{T\rho}\, \chi_k + C_{\rho \rho} \right) \\
g(\chi_k) = \frac{2}{\text{det}(C)}  \left\{ \chi_k \left[ C_{T\rho}\, \left(\overbar{T} - T_k\right) - C_{TT}\, 
\left(\overbar{\rho} - \rho_k\right) \right] + C_{T\rho} \, \left(\overbar{\rho}-\rho_k\right) - C_{\rho \rho} \left(\overbar{T} - T_k\right) \right\} \\
 h = \frac{1}{\text{det}(C)} \left[C_{TT}\, (\overbar{\rho}-\rho_k)^2-2 C_{T\rho}\, (\overbar{\rho}-\rho_k) \, 
(\overbar{T}-T_k)+ C_{\rho \rho}\, (\overbar{T}-T_k)^2 \right] \\
j(\chi_k) = \frac{g(\chi_k)}{\sqrt{8\ f(\chi_k)}}.
\end{gather}
Bimodality is defined as the appearance of two peaks in the probability distribution $P(\chi_k)$. 
To illustrate the conditions under which bimodality emerges, we consider a system whose parameters are the same as in Fig. \ref{fig:set_bd}. 
To make computations explicit, we also need to define the geometry of the system. We choose here a simplified cubic geometry, 
as depicted in Fig. \ref{fig:scheme_model}, where $\Omega = S\, l$.
This choice of parameters leads to a bifurcation in the system as depicted in Fig. \ref{fig:set_bd}. The temperature $T_b$ of reservoir $b$ plays the role of a control parameter.

\begin{figure}[H]
\centering
\includegraphics[scale = 0.35]{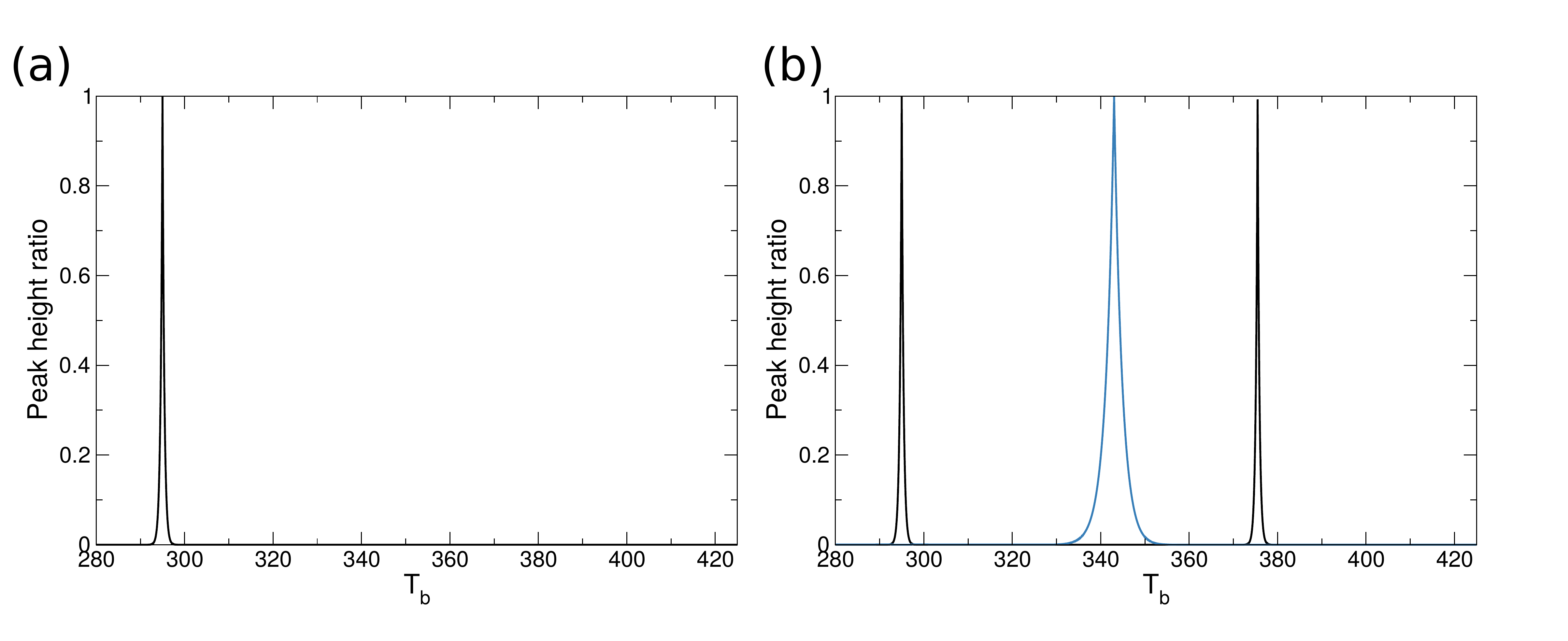}
\caption{Ratio of the probability associated with the higher peak over that of  the smaller one in the distribution $P(\chi_a)$ (a), $P(\chi_b)$ (b, black line) and 
$P(\chi_b^{-1})$ (b, blue line).
Transport and reservoirs parameters are the same as in Fig. \ref{fig:set_bd}, and $\Omega = 10^{-24}$ m$^{3}$.}
\label{fig:ratio_peaks_height}
\end{figure}

Fig. \ref{fig:ratio_peaks_height} plots the ratio of the two peaks' amplitudes for the distributions  $P(\chi_a)$ and $P(\chi_b)$ 
given by Eq. \eqref{eq:distrib_chik}, as a function of the control parameter $T_b$. Note that when the distribution is found to be unimodal, 
this ratio is set to zero. 
We observe (Fig. \ref{fig:ratio_peaks_height}(a)) that $P(\chi_a)$ is bimodal only in the vicinity of the deterministic thermal equilibrium condition, $T_a = T_b = T_m = 295 $ K. 
$P(\chi_b)$ is also bimodal under these conditions, but moreover becomes so again after the transcritical bifurcation point has been crossed, 
more precisely when the mean temperature in the system and the temperature of reservoir $b$ become equal, $T_b = \overbar{T} = 375.49$ K (see Fig. \ref{fig:ratio_peaks_height}(b)). 
More generally speaking, additional simulations reveal that bimodality is observed for a given separation efficiency $\chi_k$ whenever the system is in thermal equilibrium 
with the corresponding reservoir.   
In view of the definition of efficiency (Eq. \eqref{sep_eff}) bimodality thus also corresponds to cases where the macroscopic 
separation efficiency becomes singular as it changes sign.  
Interestingly, bimodality of the distribution is observed while the system resides in a unique steady 
state, as shown in Fig. \ref{fig:set_bd}. Unique steady states can thus give rise to bimodality in separation ratios.

\begin{figure}[H]
\centering
\includegraphics[scale = 0.62]{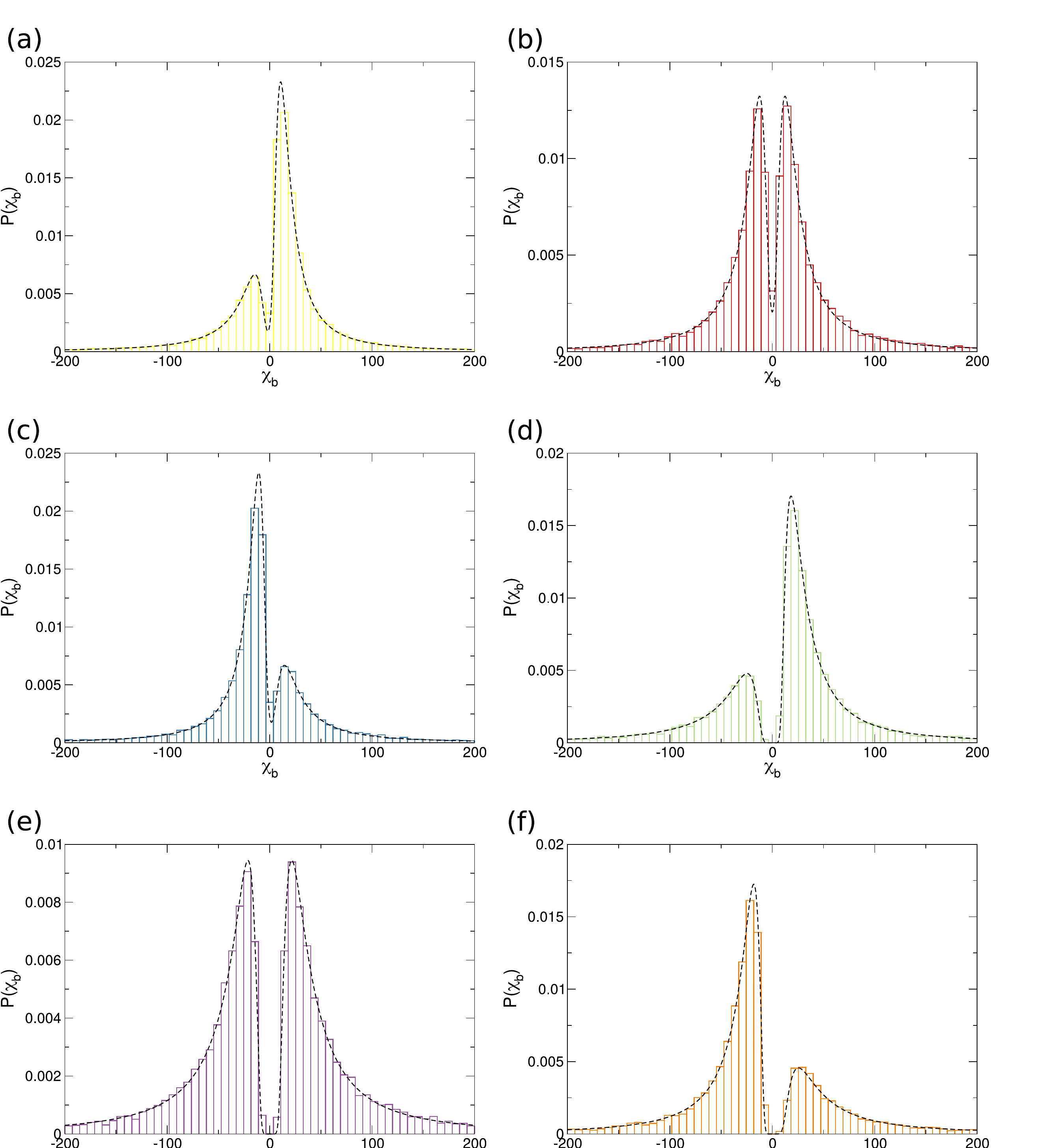}
\caption{Histograms from the numerical simulations of evolution equations \eqref{eq:SDE_C} and
\eqref{eq:SDE_T} at $T_b=294.5$ K (a), $T_b=295$ K (b), $T_b=295.5$ K (c), $T_b=373$ K (d), $T_b=374$ K (e) and $T_b=374.5$ K (f). 
The dotted lines are the analytical distributions provided by Eq. \eqref{eq:distrib_chik}. 
Parameters: transport and reservoirs parameters are the same as in Fig. \ref{fig:set_bd}, $t = 10^{-3}$s, $dt = 10^{-8}$s, $N_{real} = 20000$ 
and $\Omega = 10^{-24}$ m$^{3}$.}
\label{fig:set_distrib_chib}
\end{figure}

Numerical simulations of the full stochastic differential equations \eqref{eq:SDE_T} and \eqref{eq:SDE_C} were carried out with the Euler-Maruyama algorithm \cite{gardiner_stochastic_2009}. 
Histograms over $N_{real}$ trajectories of a duration $t$ with a time step $dt$ were produced from the simulations. They reproduce accurately the predicted emergence and disappearance of bimodality around these points, as 
shown in Fig. \ref{fig:set_distrib_chib}. It should be noted that the numerical histograms are often very well 
reproduced by the analytical expression \eqref{eq:distrib_chik}.

\subsubsection{Fluctuations and efficiency}

A question that naturally follows is how fluctuations affect the overall efficiency of the thermodiffusive process, in particular for cases where 
bimodal distributions are found. A first difficulty that arises in this context is how to choose a statistical quantity associated to the fluctuating efficiency 
that can be compared to the macroscopic efficiency. 
Indeed, the probability distribution \eqref{eq:distrib_chik}  has non-converging moments,  
which excludes the use of the mean of $\chi_k$  as a relevant quantity. This lack of convergence also appears 
with numerical simulations of the full stochastic differential equations. We will thus rather analyze how the most probable efficiency $\chi_k^{\star}$ 
compares with the macroscopic value.

\begin{figure}[H]
\centering
\includegraphics[scale = 0.70]{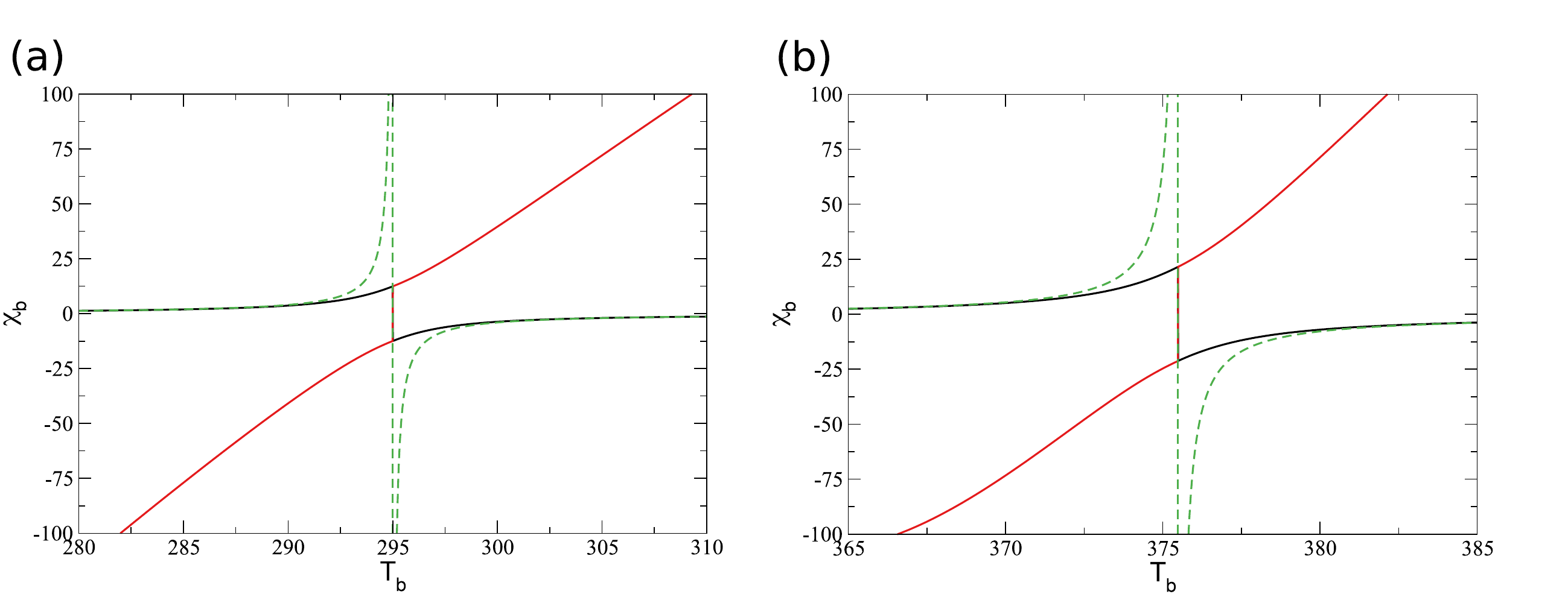}
\caption{Evolution of the most probable value $\chi_b^{\star}$ with the control parameter $T_b$ around 
$295$ K (a) and around $373.9$ K (b). 
The black line denotes the position of the highest peak of the distribution and the red line that of the 
second peak computed from Eq. \eqref{eq:distrib_chik}. The green dotted line represents the macroscopic efficiency $\chi_b^m = (\rho_m - \rho_k) \slash (T_m-T_k)$. 
Kinetic and reservoirs parameters are the same as in Fig. \ref{fig:set_bd}. }
\label{fig:evolution_peaks}
\end{figure}

\begin{figure}[H]
\centering
\includegraphics[scale=0.45]{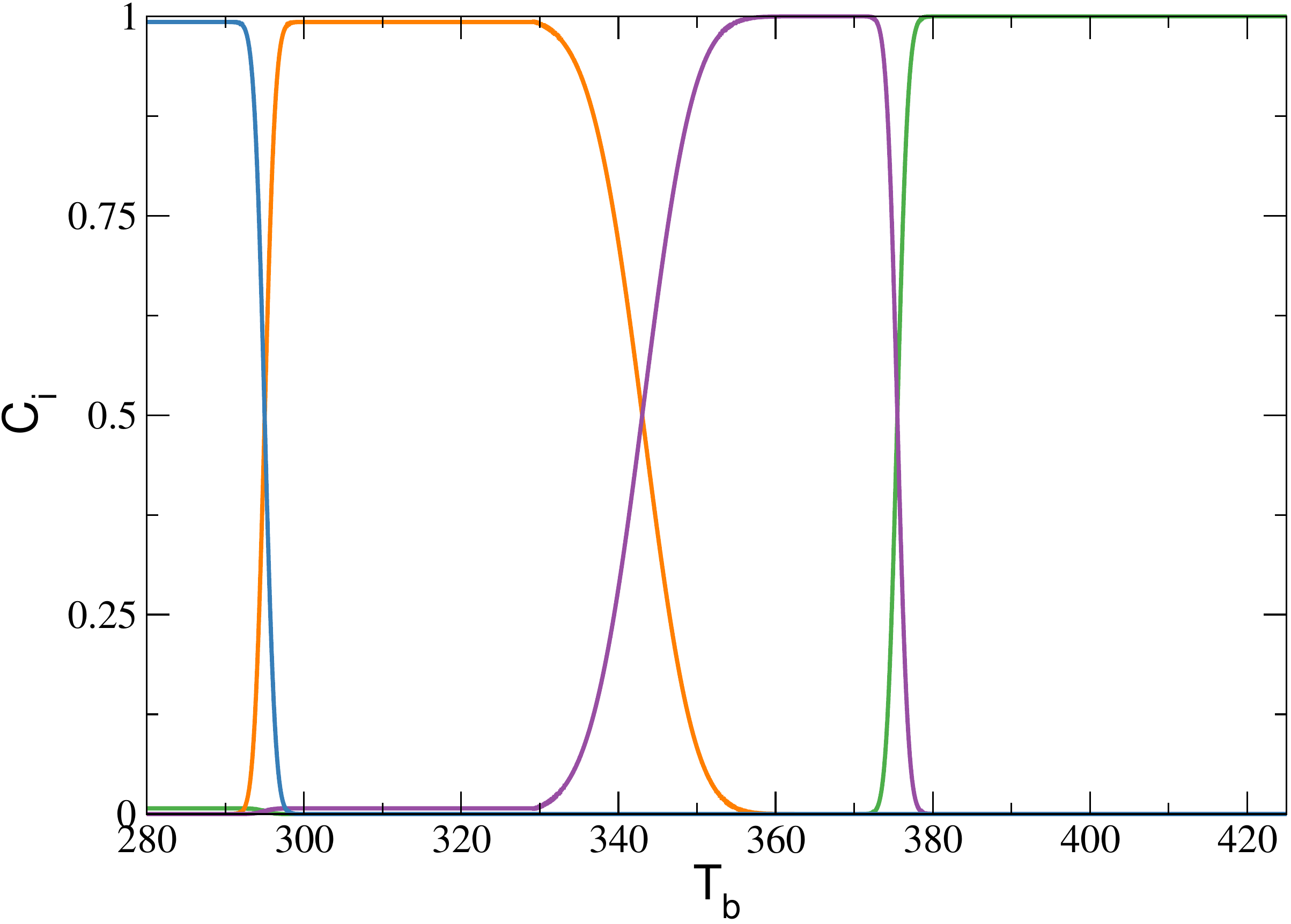}
\caption{Proportion of different separation cases $C_i$ with respect to the control parameter $T_b$. 
The first category $C_1$($T < T_b $, $\rho < \rho_b$) is denoted by the violet curve, $C_2$($T < T_b $, 
$\rho > \rho_b$) by the orange curve, $C_3$($T > T_b $, $\rho < \rho_b$) by the green curve and 
$C_4$ ($T > T_b $, $\rho > \rho_b$)  by the blue curve. Parameters: transport and reservoirs parameters are the same as in Fig. \ref{fig:set_bd}.}
\label{fig:evolution_separation_categories}
\end{figure}

Fig. \ref{fig:evolution_peaks} shows a 
comparison between the macroscopic value (green dotted curve) computed from the and the most probable value 
(black curve) of $\chi_b$ computed from the analytical expression \eqref{eq:distrib_chik}. 
We focus here on parametric regions where bimodality is found, since in other cases 
the most probable value is always very close to the macroscopic one. 
As the system goes through the points corresponding to (partial) thermal equilibria, the macroscopic 
efficiency diverges and switches from positive to negative values, as noted above.
The most probable efficiency, on the other hand, does not present such divergence. 
Its absolute value is always smaller than that of its macroscopic counterpart and the transition from
positive to negative values, although still discontinuous, is nevertheless finite. 
This result suggests that in the presence of fluctuations, 
the thermodiffusive system switches from a regime of positive efficiency to a regime of negative efficiency  in a smooth way. To confirm this, 
we decompose the possible values taken by the efficiency into  4 different subcases  (here explained for $\chi_b$): 
\begin{itemize}
 \item Case 1: $T < T_b $ and $\rho < \rho_b$ ($\chi_b$ is positive). 
 \item Case 2:  $T < T_b $ and $\rho > \rho_b$ ($\chi_b$ is negative). 
 \item Case 3: $T > T_b $ and $\rho < \rho_b$ ($\chi_b$ is negative). 
 \item Case 4: $T > T_b $ and $\rho > \rho_b$ ($\chi_b$ is positive). 
\end{itemize}
This decomposition highlights the fact that, for example, a positive separation efficiency can hide two different 
physical realities: one where density and temperature gradients are both positive, and another one where these gradients are both negative. 
The probability $C_i$ for the system to be in any of the 4 above subcases can be calculated explicitly for different $T_b$ with the analytical 
expression for the probability distribution $P(\chi_k)$, eq. \eqref{eq:distrib_chik} (see the Appendix for more details). 
The result is shown in Fig. \ref{fig:evolution_separation_categories}.
It can be seen for low $T_b$ that the  small system smoothly changes from  case 4 ($\chi_b>0$) to  case 2
($\chi_b<0$) in the vicinity of the first
thermal equilibrium, where both states become equally probable. In a sense, one could say that fluctuations allow 
the system to anticipate the appearance of a different regime of separation. As $T_b$ is further increased, the system then goes to the case 1
and finally to case 3 in a similar fashion. It should be noted that the extent of the domain over which these smooth transitions 
take place is not the same for all transitions and is moreover size-dependent, since it tends to zero as $\Omega$ increases. 

These results show that there is an intricate connection between bimodality in $\chi_k$ and transitions between the aforementioned regimes. 
The thermodiffusive system switches from a separation regime to another either when $\overbar{\rho} = \rho_k$, or when $\overbar{T} = T_k$. 
Close to these transitions, the amplitude of fluctuations is sufficient to make the system work in both regimes at the same time. 
However, this constant switching will not always lead to the appearance of two peaks in the probability distribution of separation 
efficiency. Comparing Figs. \ref{fig:evolution_separation_categories} and \ref{fig:ratio_peaks_height}(b), 
we observe that bimodality in $P(\chi_k)$ is found for some of the transitions, but not all of them. Indeed, bimodality appears in $P(\chi_b)$ only for 
transitions where the denominator of the macroscopic separation efficiency vanishes: for instance, the transition point at $T_b = 343.03$ K, 
where the numerator $\overbar{\rho} - \rho_b$ is zero (violet to orange curve), 
does not produce a bimodal distribution.
To emphasize the importance of the denominator, we can check the presence of
bimodality in the inverse of the separation ratio,
\begin{equation}
\chi_k^{-1} = \frac{T-T_k}{\rho - \rho_k}, \qquad k = a,b.
\end{equation}
Ratios of the peaks in $P(\chi_b^{-1})$ are plotted in Fig. \ref{fig:ratio_peaks_height}, showing bimodality around the 
transition point which was absent from $P(\chi_b)$. Moreover, the broad interval of $T_b$ where 
bimodality is found corresponds to the interval where the transition from case 1 to case 2 takes place. 
These differences between the distribution of the separation efficiency and its inverse were also observed in the 
distributions $P(\chi_a)$ and $P(\chi_a^{-1})$. 
In order to detect conditions under which different separation regimes can coexist in a small thermodiffusive system,
both the distribution of $\chi_k$ and of its inverse should thus be considered.

\subsubsection{Emergence of the thermodynamic limit}

We now turn to the question of the  thermodynamic limit of the system under consideration. 
The thermodynamic limit is usually defined either as the behavior of a given system in the limit of very long times, 
or as its behavior for very large systems.

\begin{figure}[H]
\centering
\includegraphics[scale = 0.40]{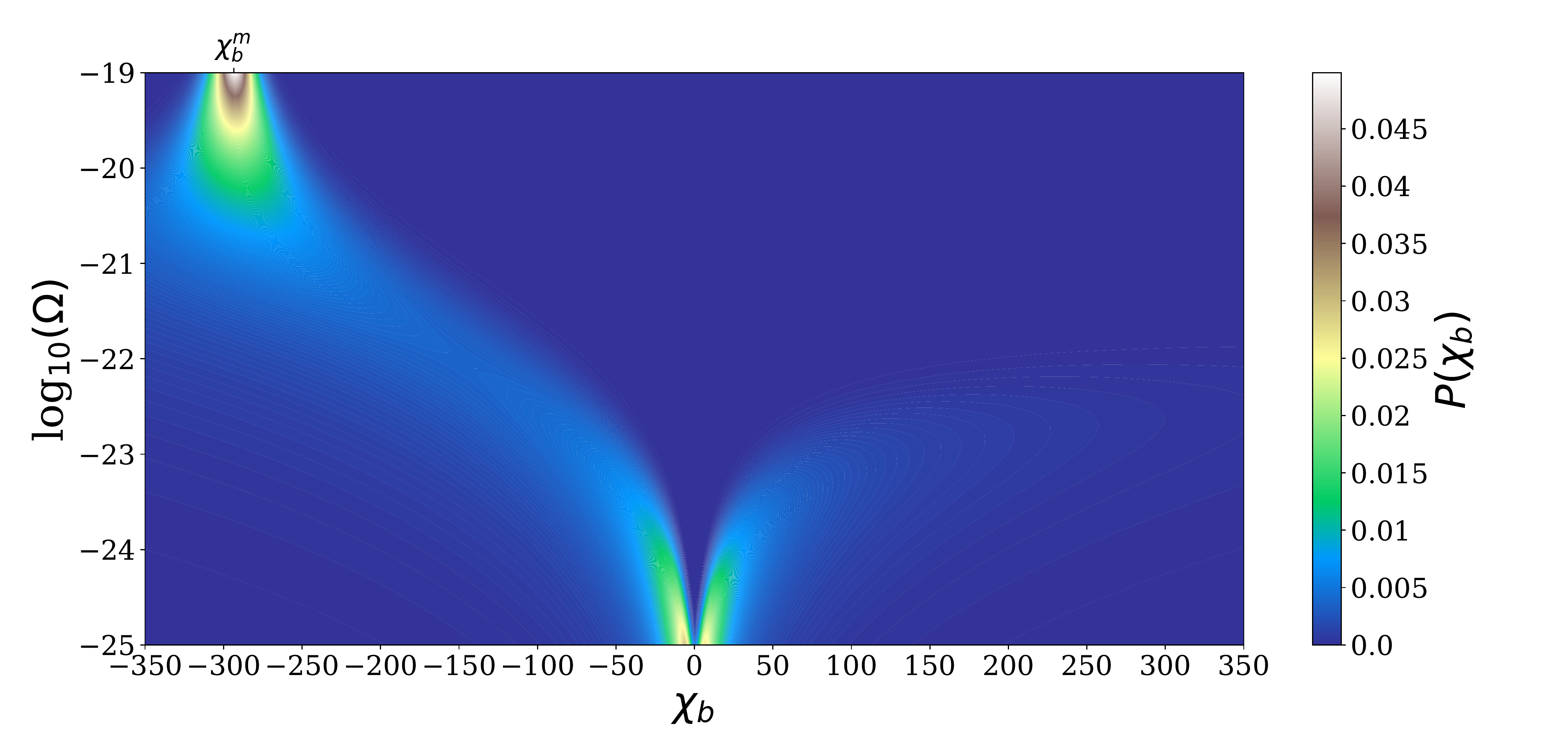}
\caption{Evolution of the distribution $P(\chi_b)$ with the size of the system $\Omega$. The parameters are the same as in Fig. \ref{fig:set_bd}, and $T_b = 375.6$ K.}
\label{fig:influence_Omega_Pchib_Tb3756}
\end{figure}

\begin{figure}[H]
\centering
\includegraphics[scale = 0.44]{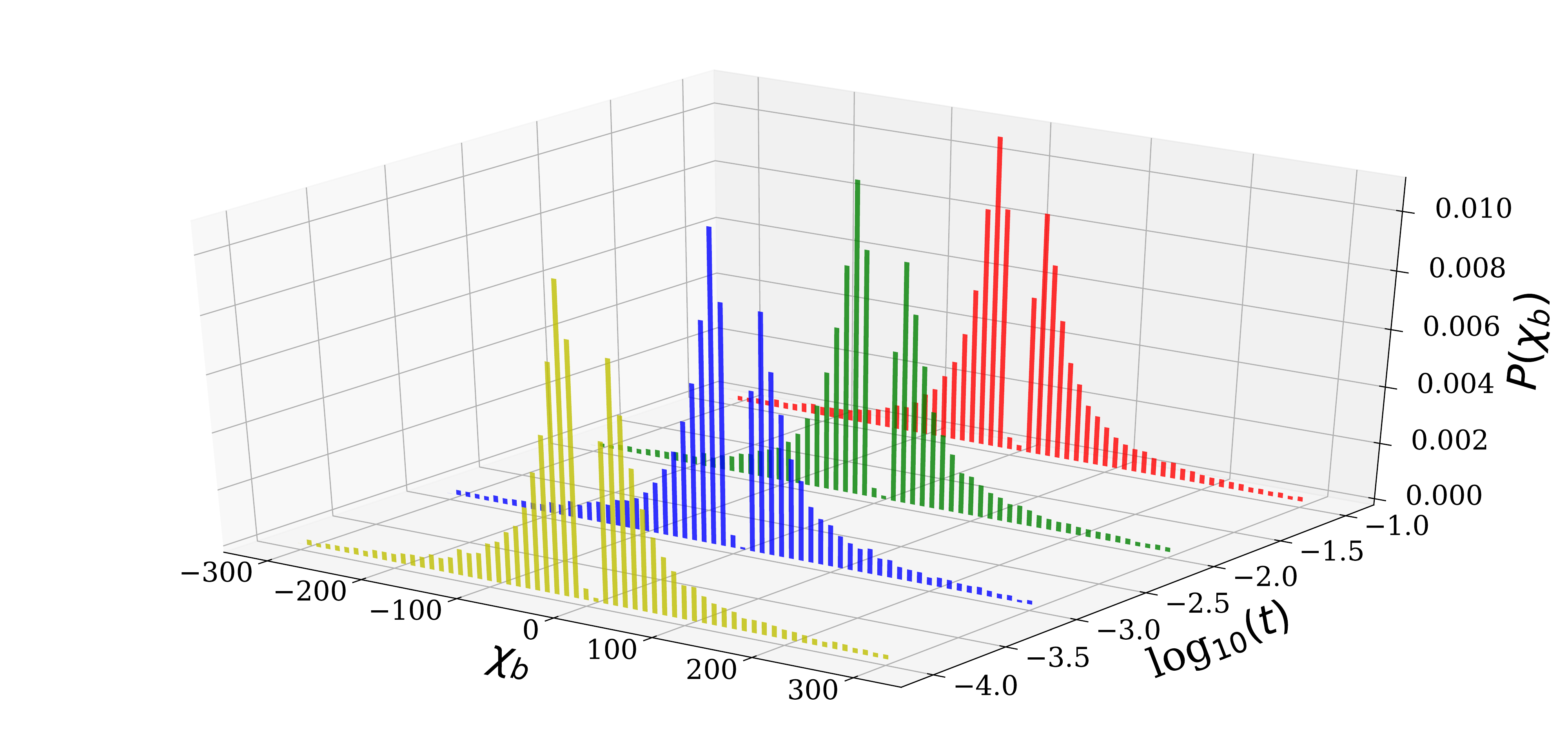}
\caption{Evolution of the distribution $P(\chi_b)$ for longer trajectories. Transport and reservoirs parameters are the same as in Fig. \ref{fig:set_bd}, 
$T_b = 375.6$ K, $dt = 10^{-8}$ s, $N_{real} = 20000$.}
\label{fig:influence_ttot_Pchib_Tb37393}
\end{figure}

In the limit of large systems, the separation efficiency tends to the expected macroscopic value as can be seen in Fig. \ref{fig:influence_Omega_Pchib_Tb3756}, 
where the analytical distribution \eqref{eq:distrib_chik} is plotted for different values of $\Omega$. We considered here a situation where 
$P(\chi_b)$ is initially bimodal. As $\Omega$ 
increases,  one of the peaks vanishes while the other peak, which is initially centered close to zero, gradually moves so that the maximum of the probability distribution 
converges towards the macroscopic value of the separation efficiency.
The evolution of $P(\chi_b)$ displays however a different behavior in the long-time limit, as shown in Fig. \ref{fig:influence_ttot_Pchib_Tb37393}: 
bimodality is conserved and the shape of the distribution is not impacted by the duration of trajectory. Two different separation efficiencies 
will be both highly probable, instead of one as in the large size limit.

The reason behind this difference lies in the way the system size and time affect the variance 
of the stochastic variables appearing in the separation efficiency. On one hand, and in view of the fluctuation-dissipation 
relations used in the stochastic approach developed in Section III, the variance of the state variables 
scales as the inverse of the system size. As $\Omega$ becomes larger,  it becomes more and more difficult for the 
system to fluctuate between the different separation regimes. The system thus spends most of its time 
in a state whose efficiency remains close to its macroscopic value.
On the other hand, since the state variables are in the weak noise limit coupled Ornstein-Uhlenbeck processes, their 
variance rapidly (exponentially) converges to a stationary value that does not depend on the duration 
of the trajectory. The probability to switch 
from a regime to another is thus not affected by the time of sampling and the bimodal shape of the probability distribution is conserved.

\subsection{Stochastic thermodynamic efficiency}

Since the thermodynamic efficiency is defined as a ratio of entropy productions, it is important to first discuss how we define stochastic entropy production.

\subsubsection{Stochastic entropy production}

In classical non-equilibrium thermodynamics, the total entropy production per unit volume (or in short, the dissipation) of a system can be expressed as the sum of  
the product of fluxes $F_i$ and associated thermodynamic forces $X_i$: 
\begin{equation}
\sigma = \frac{1}{\Omega}\, \sum_{i} \, F_i \, X_i.
\end{equation}
In the case hereby considered, $\sigma$ can also be expressed as the sum of contributions 
related to heat transfer and to matter transport for each reservoir $k$:
\begin{equation}
\sigma = \sum_{k} \left(\sigma^k_q + \sigma^k_{1}\right). 
\end{equation}
The different contributions read \cite{groot_non-equilibrium_2011} 
\begin{align} 
\sigma^k_q = \frac{F^k_q}{\Omega} \, \left( \frac{1}{T} - \frac{1}{T_k} \right)   \\ \label{sigma_1}
\sigma^k_{1} = \frac{F^k_{1}}{\Omega}\,  R_s\,  \ln \left( \frac{\rho_k}{\rho}\right),
\end{align}
where $F^k_q$ and $F_1^k$ are given by eqs. \eqref{eq:heat_fluxesa}-\eqref{eq:diffusive_fluxesb}.  
In the framework of the extended local equilibrium approach, the contributions to entropy production have the same structure, with the difference 
that the fluxes now contain a fluctuating contribution,
\begin{align} 
\sigma^k_q =  \left(\frac{F^k_q+ f^k_q}{\Omega}   \right)\,  \left( \frac{1}{T} - \frac{1}{T_k} \right)  \label{eq:sigma_k_q} \\
\sigma^k_{1} =  \left( \frac{F^k_1+ f^k_1}{\Omega}   \right) \, R_s\, \ln{\left( \frac{\rho_k}{\rho}\right)} \label{eq:sigma_k_rho}
\end{align}
Consequently, each contribution to $\sigma$ 
can be split into a systematic part $\left(\sigma_{i}^k\right)_{sys}$
\begin{equation}
\left(\sigma_{i}^k\right)_{sys} = \frac{F_i^k}{\Omega} \, X_i^k
\end{equation} 
and a part arising from the effect of fluctuations $\left(\sigma_{i}^k\right)_{fluct}$
\begin{equation}
\left(\sigma_{i}^k\right)_{fluct} = \frac{f_i^k}{\Omega} \, X_i^k.
\end{equation}

At the macroscopic scale, the thermodynamic efficiency $\eta_k$ associated to thermodiffusion with respect to a given reservoir $k$ 
is defined as the ratio of instantaneous values of the different contributions to entropy productions (see eq. \eqref{thermo_eff}).
In the stochastic approach that we adopt, these contributions are proportional to Gaussian white noises,  which could result in the 
appearance of singularities. To avoid such situations, we will consider in what follows the time-averaged entropy productions  
\begin{equation}
\Sigma^k_i = \frac{1}{t} \int_{0}^t \, \sigma^k_i(t^{\prime})\,  dt^{\prime}. 
\end{equation}
Notice that 
the noise present in the different contributions $\sigma^k_i$ to the stochastic entropy production is not additive but multiplicative. 
This makes the derivation of general analytical expressions
for $\sigma_i^k$ or for  its time-averaged counterpart $\Sigma_i^k$ a rather difficult task. 
However,  $\Sigma_i^k$ can be shown to obey a 
Gaussian distribution at the steady state and in the weak-noise limit. Indeed, developing the dissipation $\sigma$ around its deterministic non-equilibrium stationary state 
$\overline{\sigma}$, we obtain to the leading order in $\Omega^{-1}$:
\begin{equation}
\delta \Sigma_{i}^{k} = \Sigma_i^k - \overline{\Sigma}_i^k = \frac{1}{t} \mathop{\int_0^t} 
\left( \left.\frac{ \partial \left(\sigma_{i}^k\right)_{sys} }{\partial T} \right\rvert_{\overbar{T}, \overbar{\rho}}\ \delta T 
+ \left. \frac{\partial \left(\sigma_{i}^k\right)_{sys}}{\partial \rho}\right\rvert_{\overbar{T}, \overbar{\rho}}\  \delta \rho + 
\frac{f_i^k}{\Omega} \, X_i^k \left(\overbar{T}, \overbar{\rho} \right) \right) dt^{\prime}.
\label{eq:expansion_sigma}
\end{equation}
The two first terms are integrals of Ornstein-Uhlenbeck processes and  define zero-mean normally distributed variables in the long-time limit. 
The last term is the time average of a Wiener process, which is also normally distributed and has a mean of zero. The sum of all these contributions 
will thus also be a Gaussian variable (see \cite{de_decker_stochastic_2016} for more details). 

\begin{figure}[H]
\centering
\includegraphics[scale = 0.50]{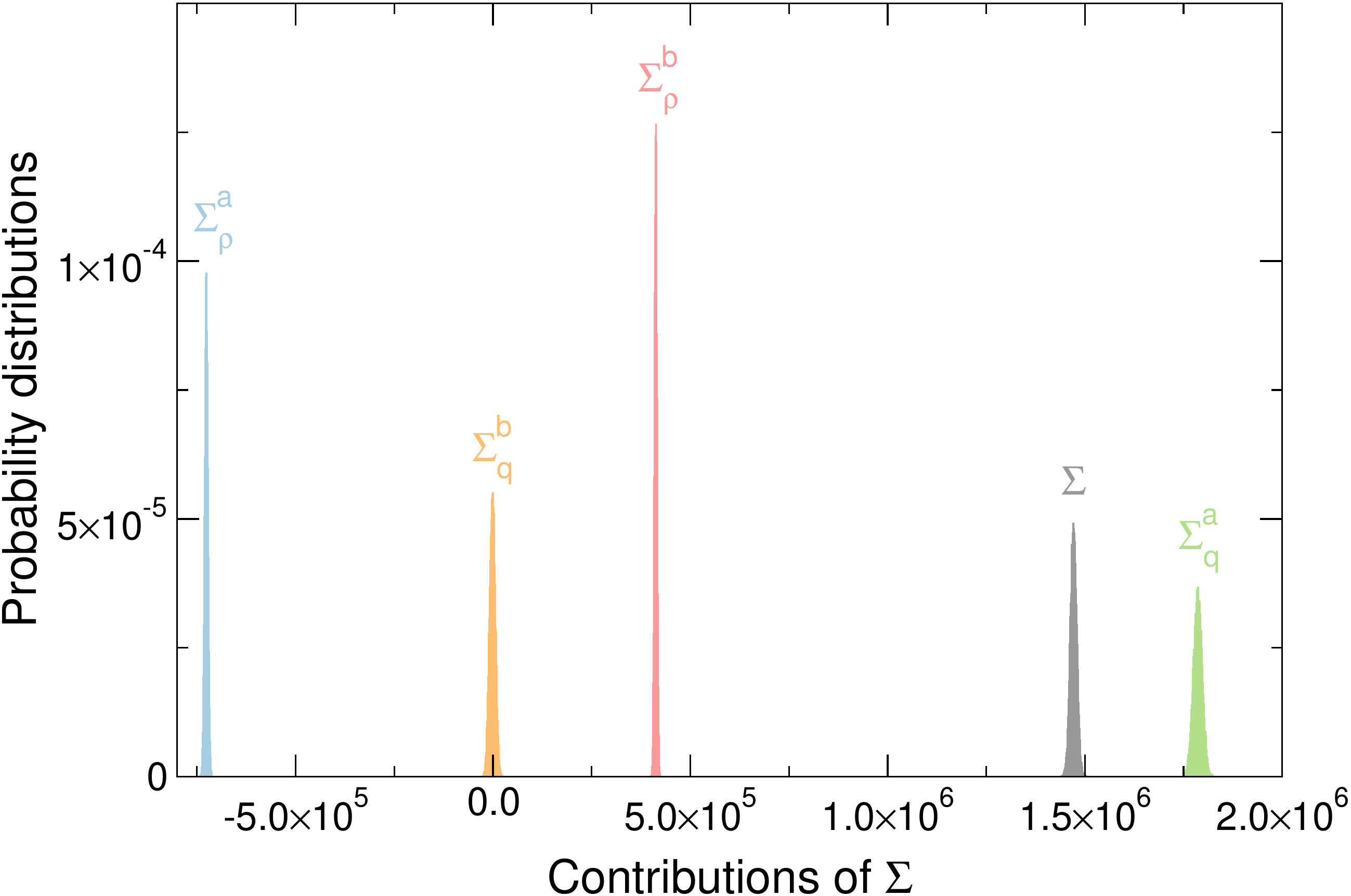}
\caption{Histograms of the different contributions to $\Sigma$  resulting from numerical 
simulations of stochastic trajectories. Parameters are 
the same as in Fig. \ref{fig:set_bd}, exepct for $T_b = 375.6$ K, $t = 10^{-3}$ s, $dt = 10^{-8}$s, $N_{real} = 20000$.}
\label{fig:contributions_dissipations_Tb3756}
\end{figure}

\begin{figure}[H]
\centering
\includegraphics[scale = 0.70]{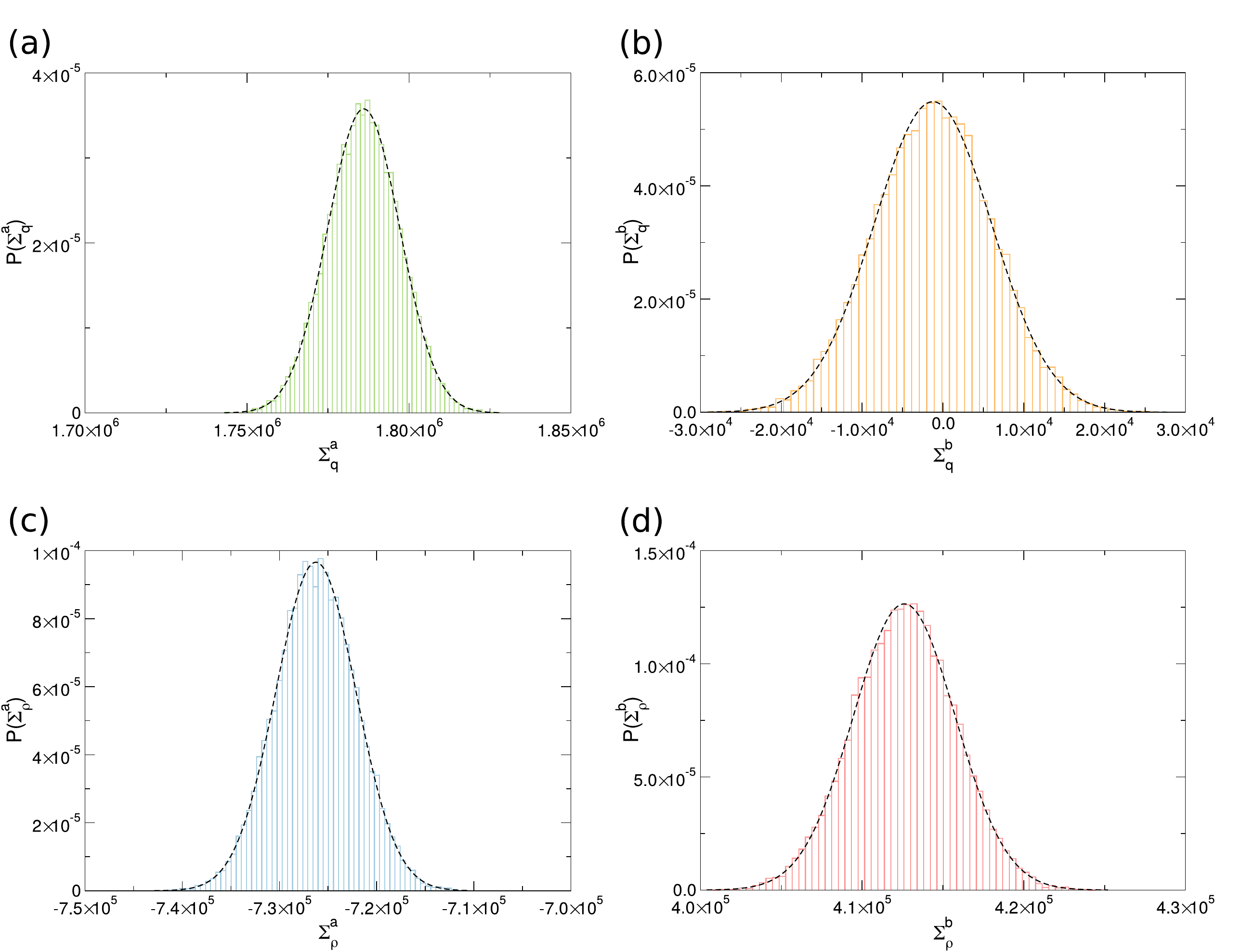}
\caption{Probability distribution of each contribution to $\Sigma$ resulting from numerical 
simulations of stochastic trajectories. The dotted lines are Gaussian distributions whose mean and variance come from 
the numerical simulations. 
The parameters are the same as in  Fig. \ref{fig:contributions_dissipations_Tb3756}. }
\label{fig:contributions_dissipations_Sigma_i^k}
\end{figure}

As an illustration of this, 
the histograms of the different contributions to 
\begin{eqnarray}
\Sigma = \sum_{k} \left(\Sigma^k_q + \Sigma^k_{1}\right) = \Sigma_q^a + \Sigma_q^b +  \Sigma_1^a + \Sigma_1^b
\end{eqnarray}
are depicted in Fig. \ref{fig:contributions_dissipations_Tb3756}
for a given value of $T_b$. Despite the small size of the system considered in this example, each contribution defines 
an approximately Gaussian distribution centered around a well-defined mean, as can also be seen in Fig. \ref{fig:contributions_dissipations_Sigma_i^k}. 
Some  contributions are centered around negative values ($\Sigma_{1}^a$), 
others around positive values ($\Sigma_{q}^a$) and one of them remains close to zero ($\Sigma_{q}^b$).

\subsubsection{Probability distribution  of the thermodynamic efficiency}

We consider in this Section the statistical properties of the stochastic thermodynamic efficiency, which we thus need to define in more detail. 
As previously stated, the macroscopic thermodynamic efficiency is usually defined as the ratio of the dissipation of the 
non-spontaneous process over the dissipation of the spontaneous one. This means that 
the very definition of $\eta_k$ can change as the parameters are varied, since the system can switch from a regime to another (for example, 
from a heat pump regime to  
a matter pump regime). 
To avoid ambiguities in defining thermodynamic efficiency for each situation, we choose to denote by $\eta_k$
the ratio of mass transport dissipation over heat transport dissipation: 
\begin{equation}
\eta_k = - \frac{\Sigma^k_{1}}{\Sigma^k_q}.
\end{equation}
As mentioned above, we rely on time-averaged dissipations along a stochastic trajectory of duration $t$:
\begin{align}
\Sigma^k_q = \frac{1}{t} \int_{0}^t \sigma^k_q(t^{\prime}) dt^{\prime} \\
\Sigma^k_{1} = \frac{1}{t} \int_{0}^t \sigma^k_{1}(t^{\prime}) dt^{\prime}
\end{align}
to avoid singularities. 
Note that we will be interested in the statistical properties of both $\eta_k$ and $\eta_k^{-1}$

The fact that $\Sigma^k_{1}$ and $\Sigma^k_{q}$ are distributed in an almost Gaussian fashion suggests that $\eta_k$ and/or $\eta_k^{-1}$, just like the 
separation efficiency $\chi_k$, could obey non-trivial probability distributions including bimodality. 
However, the lack of analytical results prevents  using the same method as in Section \ref{subsec:stochastic_separation_efficiency}
to detect the emergence of such bimodal distributions. We will instead rest to do so on the ``dip test'', which was  originally 
proposed in \cite{hartigan_dip_1985}.

\begin{figure}[H]
\centering
\includegraphics[scale = 0.75]{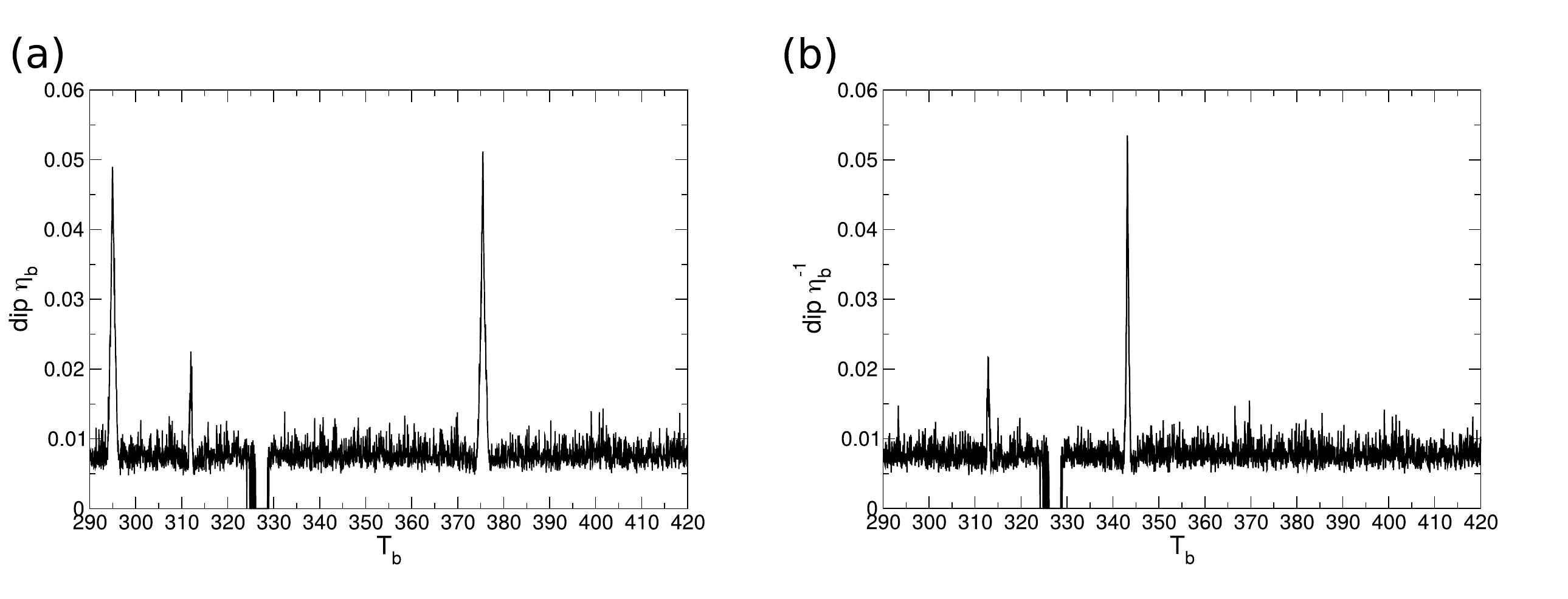}
\caption{Dip computed for the numerical distributions $P(\eta_b)$ (a) and $P(\eta_b^{-1})$ (b) 
out of 1000 realizations. 
Transport and reservoirs parameters are the same as in Fig. \ref{fig:set_bd}, with $t = 10^{-2}$ s, $dt = 10^{-8}$ s and $N_{real} = 1 000$.}
\label{fig:dip}
\end{figure}

The dip is a statistical indicator measuring the departure of an empirical distribution from unimodality. 
The computed dip is the maximum difference between the empirical cumulative distribution function and the unimodal 
cumulative distribution function which minimizes that difference. When the distribution is unimodal, the dip vanishes 
for exponentially decreasing probability densities and tends to a small positive constant otherwise. 
When multimodality is present in the probability density, the dip increases significantly.
Fig. \ref{fig:dip} displays the computation of the dip in $P(\eta_b)$ and $P\left(\eta_b^{-1}\right)$ for 1000 realizations of 
stochastic trajectories at different values of $T_b$. Notice that the weak fluctuations of the dip result from the finite size of the sampling. 
As previously observed in the analysis of the separation efficiency, multimodality appears for well-defined values 
of the control parameter. 
Multimodality appears in $P(\eta_b)$ at $T_b = 294.95$ K, $T_b = 312.00$ K and $T_b = 375.5$ K, and for $P\left(\eta_b^{-1}\right)$ at $T_b = 312.85$ K and $T_b = 343.15$ K.
Computations of the histograms at those points confirm that these multimodalities indeed correspond to bimodalities. 

In a way similar to what was observed for $\chi_k$, the emergence of bimodality can be correlated to situations where the mean 
of the denominator of $\eta_k$ (or $\eta_k^{-1}$) vanishes. 
At the macroscopic scale, situations where the denominator of either $\eta_k$ or its inverse vanish would correspond to changes 
in the spontaneity of one of the processes, since the entropy production associated to it changes sign for these conditions.
For example, the bimodality appearing in $P(\eta_b)$ 
at $T_b = 294.95$ K and $T_b = 375.5$ K corresponds to cases where $\overbar{T} = T_b$ so that to the leading order in 
$\Omega^{-1}$, $\overline{\Sigma_q^b}= 0$ since the mean thermodynamic force is zero (see eq. \eqref{eq:sigma_k_q}).
At these points, both $\chi_k$ and $\eta_k$ thus obey bimodal distributions.  
However, while for the separation efficiency a vanishing mean of the denominator can only be achieved for $\overbar{T} = T_k$, 
more possibilities exist for cancelling the mean of the entropy production appearing in the thermodynamic efficiency. 
Indeed, following eq. \eqref{eq:sigma_k_q}, the mean entropy production due to heat dissipation also vanishes (again, to the leading order in $\Omega^{-1}$) 
whenever the mean thermodynamic flux becomes zero. Predicting the parametric conditions under which this 
happens is not trivial, but the cancelling of the fluxes can be observed numerically. For $\eta_b$, this leads to the appearance of an 
additional region of bimodality around $T_b = 312.00$ K. A similar situation is observed for the distribution of $\eta_b^{-1}$: the peak at  
$T_b = 343.15$ K corresponds to a vanishing mean force, and the peak at  $T_b = 312.85$ K corresponds to a vanishing mean flux. 
To emphasize the importance of the value on the denominator, note that for a same value of control parameters, 
the distribution of an efficiency will be bimodal but its inverse will not, even though the physical situation is identical.

\subsubsection{Size and sampling time effects on the distribution of thermodynamic efficiency}

\begin{figure}[H]
\centering
\includegraphics[scale = 0.33]{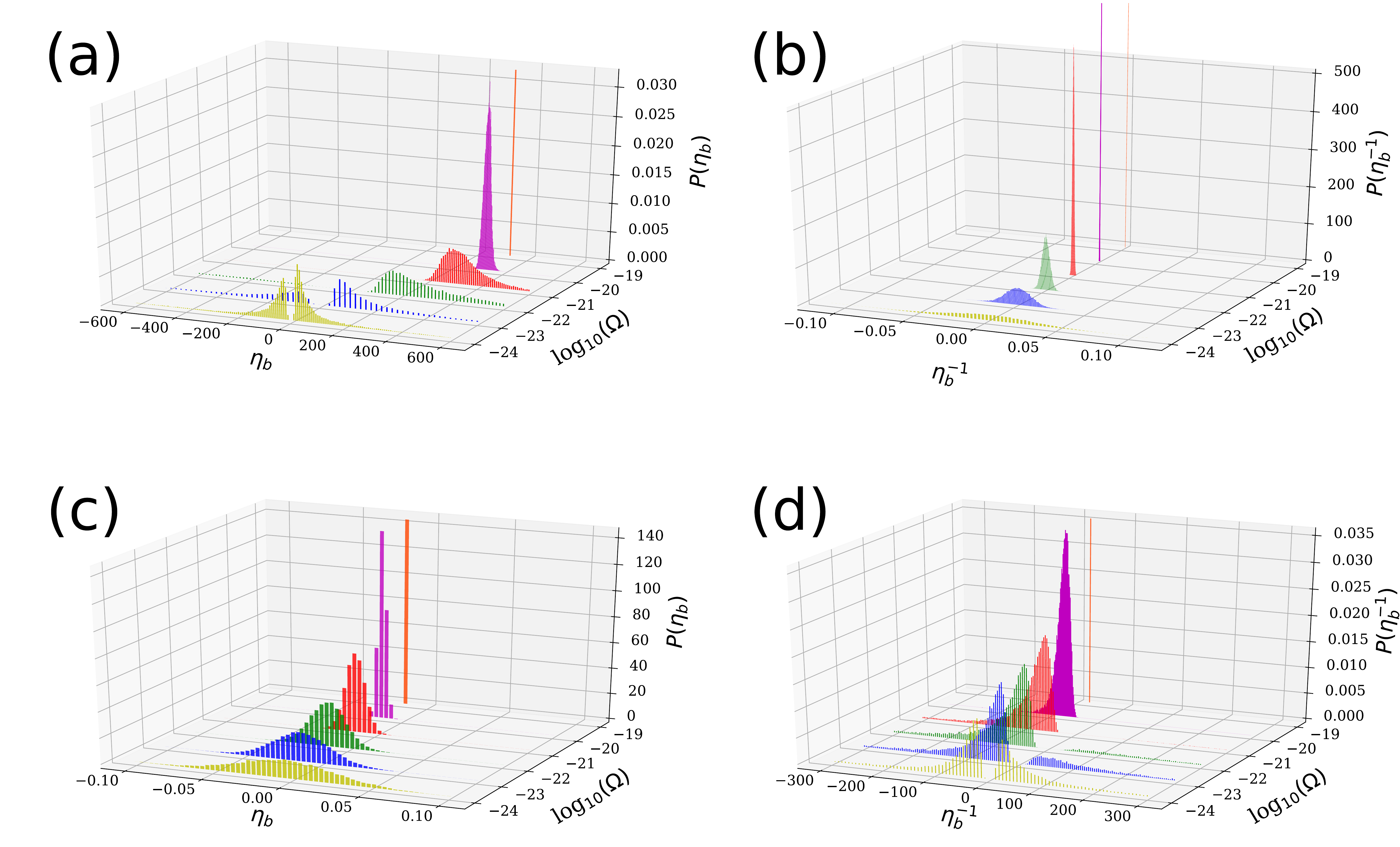}
\caption{Influence of the system size on the probability distribution $P\left(\eta_b\right)$ and $P\left(\eta_b^{-1}\right)$ at $T_b = 375.6$ K ((a) 
and (b)) and at $T_b = 343.31$ K ((c) and (d)). Orange bars denote the 
macroscopic values $\eta_b^m$ and $(\eta_b^m)^{-1}$. At $T_b = 375.6$ K, $\eta_b^m = 334.02$ and $(\eta_b^m)^{-1} = 0.0030$. At $T_b = 343.31$ K, 
$\eta_b^m = -0.014$ and $(\eta_b^m)^{-1} = -68.91$. Transport and reservoirs parameters are the same as in Fig. \ref{fig:set_bd}, $t = 10^{-4}$ s, $dt = 10^{-8}$ s, $N_{real} = 20000$.}
\label{fig:eta_b_inv_eta_b_Omega}
\end{figure}

Despite the absence of analytical expressions for $P(\eta_b)$ and $P(\eta_b^{-1})$, 
the large size and long time limits can be investigated through numerical simulations. 
Both distributions were computed at temperatures $T_b = 375.6$ K and $T_b = 343.31$ K, where bimodality was observed for short times and small systems.
At $T_b = 375.6$ K, the originally bimodal distribution $P(\eta_b)$ converges in the limit of large systems to a unimodal distribution whose most probable value is the macroscopic value (see Fig. \ref{fig:eta_b_inv_eta_b_Omega}). Moreover, the amplitude of fluctuations around this mean decreases as the system becomes larger. The originally unimodal distribution $P(\eta_b^{-1})$ follows the same tendency, with a decreasing variability and a convergence to the macroscopic value.
Similar conclusions can be made for $T_b = 343.31$ K, where $P(\eta_b^{-1})$ was observed to be bimodal.  
Even though the most probable value first peaked on the right of the distribution in small systems, the peak converges to the macroscopic value in macroscopic systems.
The situation is here analogous to what was observed for the effect of the system size on the distribution of the separation efficiencies. 

\begin{figure}[H]
\centering
\includegraphics[scale = 0.28]{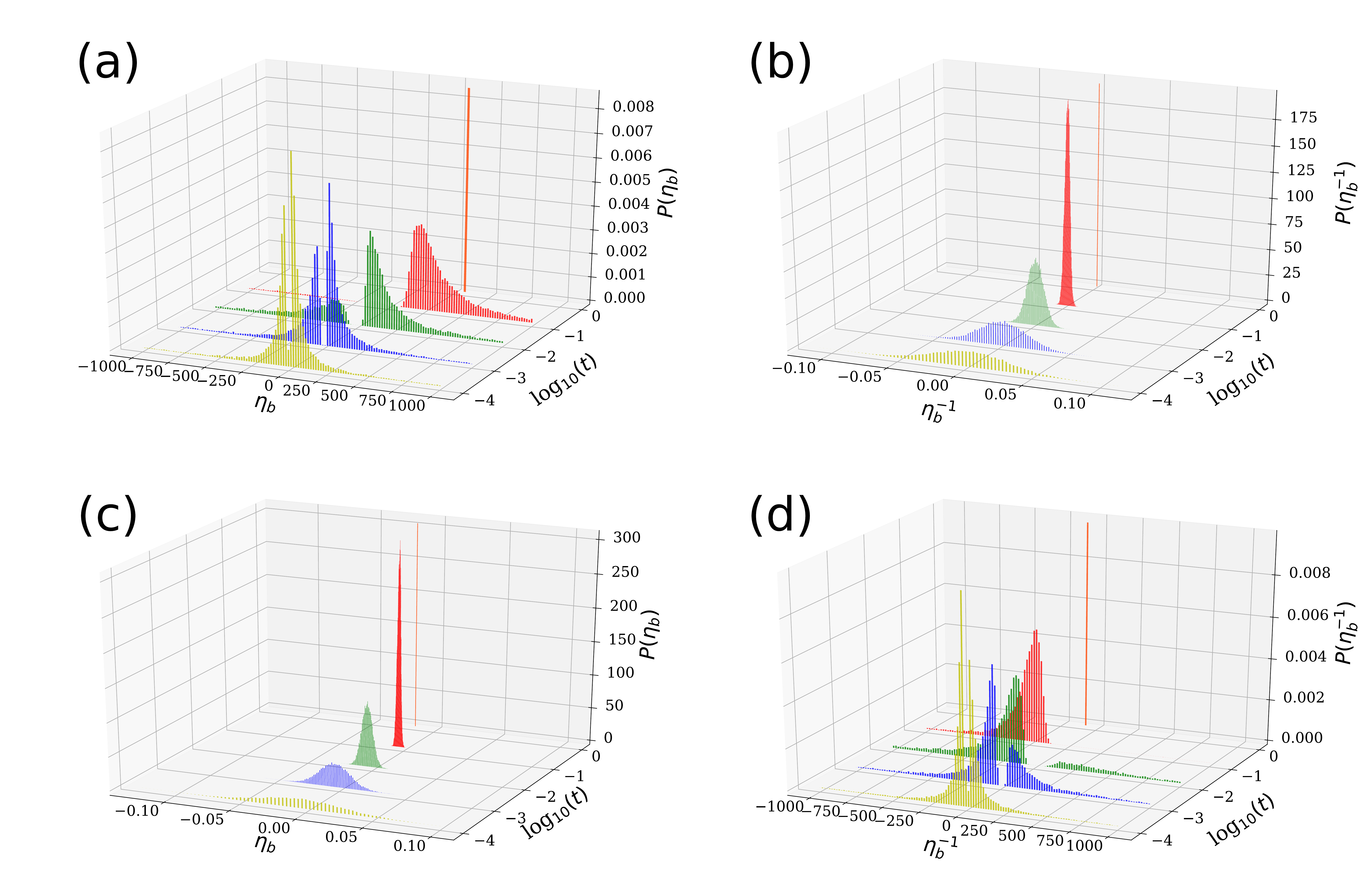}
\caption{Influence of the sampling time $t$  on the probability distribution $P\left(\eta_b\right)$ and $P\left(\eta_b^{-1}\right)$ at $T_b = 375.6$ K
((a) and (b)) and at $T_b = 343.31$ K ((c) and (d)).
Orange bars denote the macroscopic values $\eta_b^m$ and $(\eta_b^m)^{-1}$. At $T_b = 343.31$K, $\eta_b^m = -0.014$ and $(\eta_b^m)^{-1} = -68.91$.
Transport and reservoirs parameters are the same as in Fig. \ref{fig:set_bd}, with $dt = 10^{-8}$ s and $N_{real} = 20 000$.}
\label{fig:eta_b_inv_eta_b_ttot}
\end{figure}

The long-time limit shows interesting departures from the distributions taken at large size limit. In both cases bimodality,
when present, tends to disappear in favor of a unimodal distribution (see Fig. \ref{fig:eta_b_inv_eta_b_ttot}). The amplitude of fluctuations also decreases, but 
contrary to the large system limit the most probable value for large times does not converge to the macroscopic value but to an 
entirely different one, having sometimes an opposite sign. This shift can be attributed to the non-linearities present in the 
stochastic entropy production. In the limit of large systems high order moments such as the average of $T^k$ or $\rho^k$ 
tend to factorize as $\overline{T}^k$ and $\overline{\rho}^k$, respectively, since the mean deviations $\delta T^k$ and $\delta \rho^k$ 
scale as the inverse of the system size. This factorization is not expected to hold for small systems, even in the limit of long times. 

Remember that for the separation efficiency, bimodality is preserved even for long sampling times. This is no more the case for the thermodynamic efficiency, for which the bimodality is only transient.
This different behavior can be rationalized on the basis of the time dependence of the variances of the variables entering the
ratios that define these two efficiencies. 
Thermodynamic efficiencies are not ratios of Ornstein-Uhlenbeck processes, as is the case for the separation efficiency, but ratios of 
\textit{integrals} of Ornstein-Uhlenbeck and Wiener processes (at least, in the weak noise limit). The main difference between these two situations 
lies in the temporal dependence of the variances. A stationary Ornstein-Uhlenbeck 
process is characterized by a constant variance, while that of its integral and of the integral of a Wiener process are linear in time. 
So in the long time limit the mean of the dissipations will remain constant but their standard deviations (STD) will decrease as the square root 
of $t$, since the integrated dissipations are divided by $t$. 
The integral does not affect, however, the size dependence of the different moments of the dissipations. 
The mean dissipation is an intensive quantity and the dominant contributions to the variance (as computed from Eq. \eqref{eq:expansion_sigma}) scale as $\Omega^{-1}$.
Consequently the coefficient of variation (CV) of the integrated dissipations, 
\begin{equation}
\text{CV}(\Sigma_i^k) = \frac{\text{STD}(\Sigma_i^k)}{\overbar{\Sigma_i^k}}\propto \frac{1}{\Omega \sqrt{t}},
\end{equation}
decreases with time and system size, as previously mentioned. In long time limit, this means 
that the fluctuations around the mean dissipations will be sufficiently small to prevent the system from 
switching from one regime to another. The distribution will then become unimodal. 
The time needed to observe a unimodal distribution depends on the amplitude of 
fluctuations and thus  varies with the size of the system.

\section{Conclusions}
\label{Conclusions}

In this work, we relied on a stochastic extension of the local equilibrium hypothesis to study the efficiency of thermodiffusion 
in small compartmentalized systems.  
Starting from stochastic differential equations 
for the mass density and for the temperature of the system,   
two different measures of the efficiency of thermodiffusion were analyzed: the separation efficiency, which is a ratio between state variables and the thermodynamic efficiency,
which is a ratio between time-averaged entropy productions. 
In most cases, the probability distributions 
associated with these quantities are Gaussian and centered around their macroscopic value. 
However, both of them were seen to obey bimodal distributions 
for specific values of the control parameters whenever small systems and short times are  considered. 
These bimodalities were shown to correspond to situations where the system constantly switches between two different 
regimes, including for example transitions  between a heat pump and a matter pump regimes.

This phenomenon could be traced back to situations   
 where the mean of the denominators appearing in these separations vanishes. In the
case of separation efficiency, bimodality appears when the mean temperature of the central cell equals that of one of the reservoirs,
i.e., when a partial thermal equilibrium is established. In the case 
of thermodynamic efficiency, bimodality coincides with situations where the mean dissipation associated to one 
of the thermodynamic fluxes vanishes. This can happen either because the 
thermodynamic driving forces becomes zero or because the flux itself vanishes.

Another interesting property emerging from numerical simulations is the fact that the limit of large systems 
and the limit of long times are not equivalent. For both efficiencies, the distribution becomes unimodal 
and peaked around the macroscopic value in the large size limit. 
In the limit of long times, bimodality is conserved for the separation efficiency while it 
disappears gradually for the thermodynamic efficiency. This can be explained by the fact that the variance of the average entropy production decreases with time, while that of the state variables does not. 
Identical macroscopic constraints and stationary distributions of state variables can thus
lead to different dynamics for the distribution of an efficiency, depending on how it is defined.

While the present work focuses on thermodiffusion, a similar approach could be used to derive stochastic models for  other systems where dissipative 
phenomena are coupled, such as thermoelectric devices or active membrane transport. Other aspects of coupled transport phenomena 
like the trade-off between power and efficiency could be studied with our extension of the local equilibrium hypothesis. It would be particularly interesting to 
find relations between this trade-off and the various transport coefficients and reservoir parameters, a task for which our formalism is well suited.
Beyond the application to coupled transport phenomena, the results about the behavior of stochastic ratios are expected to be quite general because they are solely based 
on the local equilibrium assumption and should thus apply to a broader range of problems. As long as the state variables describing a system can be described by 
Ornstein-Uhlenbeck processes, ratios of these quantities 
should be expected to obey bimodal distributions whenever the mean of the denominator vanishes. This could be observed, for example, 
when computing the yield of a  process admitting several pathways. Similarly, under these conditions 
the extended local equilibrium hypothesis has the consequence  that entropy production can always be expanded as a sum of integrals of Ornstein-Uhlenbeck and Wiener processes. Ratios of dissipations, including for example the ratio of mechanical power and heat flux used in Carnot cycles or the ratio of electrical current and heat flux in fuel cells,
will thus present features that are similar to those observed here.

\section{Acknowledgements}
J.-F. D. acknowledges the F.R.S.-FNRS (Fonds National de la Recherche Scientifique) for financial support through the grant of a FRIA fellowship.

\section*{Appendix: Probabilities of separation categories}
\label{Appendix_B}
In this section the computation of the probability for each separation case, $C_i$, will be discussed explicitly. 
Suppose that we look for the probability distribution $P(y)$ of a generic ratio $y = \frac{x_1}{x_2}$, having only 
the joint probability distribution $P_{j}(x_1,x_2)$. After a simple change of variables, the probability distribution of the 
ratio can be split in two parts following the sign of the denominator $x_2$ :
\begin{align}
P^{+}(y) = \int_{0}^{+ \infty} P_j(yx_2,x_2) x_2 dx_2 \\
P^{-}(y) = \int_{0}^{- \infty} P_j(yx_2,x_2) x_2 dx_2
\end{align}
Using the same reasoning as in \cite{polettini_efficiency_2015}, an analytical expression for those two contributions
can be computed if the joint probability $P_j(x_1,x_2)$ is a multivariate Gaussian distribution.

Considering the particular case of the separation efficiency $\chi_k$, the joint probability distribution of 
$P(T-T_k,\rho-\rho_k)$ will be given by the same expression as the one derived in Section \ref{sec:stochastic_model}. 
Indeed, even if the probability distribution were computed for the fluctuations of state variables around their means, 
a simple linear variable shift enables us to find the same distribution with the same covariance matrix.

The distributions following the sign of the discrete gradient $T-T_k$ can be then written: 
\begin{align}
P^{+}(\chi_k) = \frac{1}{2 \pi \sqrt{\text{det}(M)} f(\chi_k)}\, \e^{-\frac{ h}{2}} \left[ 1- \sqrt{\pi}\,
j(\chi_k) \, \e^{ j(\chi_k)^2}\,  \text{erfc} \left( j(\chi_k) \right)\right] \\
P^{-}(\chi_k) = \frac{1}{2 \pi \sqrt{\text{det}(M)} f(\chi_k)}\, \e^{-\frac{ h}{2}} \left[ 1+ \sqrt{\pi}\,
j(\chi_k) \, \e^{ j(\chi_k)^2}\,  \text{erfc} \left( - j(\chi_k) \right)\right]
\end{align}
The probability to observe a given separation category is finally obtained by computing the proper integral 
on the chosen domain of the efficiency
\begin{align}
C_1 = \int_{0}^{\infty} P^{-}(\chi_k) d \chi_k  \qquad T < T_k, \rho < \rho_k, \chi_k >0 \\
C_2 = \int_{-\infty}^{0} P^{-}(\chi_k) d \chi_k  \qquad T < T_k, \rho > \rho_k, \chi_k < 0 \\
C_3 = \int_{-\infty}^{0} P^{+}(\chi_k) d \chi_k  \qquad T > T_k, \rho < \rho_k, \chi_k <0 \\
C_4 = \int_{0}^{\infty} P^{+}(\chi_k) d \chi_k  \qquad T > T_k, \rho > \rho_k, \chi_k >0. 
\end{align}
Those probabilities were computed for each value of the parameter $T_b$, as depicted in Fig. \ref{fig:evolution_separation_categories}.
\bibliographystyle{apsrev4-1}
\bibliography{ref_stoeff}

\end{document}